%% file: main.tex
\documentclass[10pt,twocolumn,letterpaper]{article}

\usepackage{iccv}              


\usepackage[accsupp]{axessibility}
\definecolor{iccvblue}{rgb}{0.21,0.49,0.74}
\usepackage[pagebackref,breaklinks,colorlinks,allcolors=iccvblue]{hyperref}

\usepackage{threeparttable}

\newcommand{\boldred}[1]{\textbf{\textcolor{red}{#1}}}
\newcommand{\blue}[1]{\textcolor{blue}{#1}}

\newcommand{\ourmethod}{\texttt{FED-PRIME}}

\usepackage{booktabs}
\usepackage{multirow}
\usepackage{graphicx}
\usepackage{amsmath, amssymb}
\usepackage{algorithm}
\usepackage{algpseudocode}

\def\paperID{12364} 
\def\confName{ICCV}
\def\confYear{2025}

\title{Federated Prompt-Tuning with Heterogeneous and Incomplete\\ Multimodal Client Data}

\author{
Thu Hang Phung \textsuperscript{1} \quad 
Duong M. Nguyen \textsuperscript{1} \quad 
Thanh Trung Huynh \textsuperscript{2} \quad 
Quoc Viet Hung Nguyen \textsuperscript{3} \quad \\
Trong Nghia Hoang \textsuperscript{4}\thanks{Corresponding Authors} \quad 
Phi Le Nguyen \textsuperscript{1}\footnotemark[1] \quad  \\
\textsuperscript{1} Institute of AI Innovation and Societal Impact, Hanoi University of Science and Technology \\ \textsuperscript{2} EPFL \quad \textsuperscript{3} Griffith University \quad \textsuperscript{4} Washington State University \\
}

\begin{document}
\maketitle

\newcommand{\phile}[1]{\textcolor{red}{#1}}

\begin{abstract}
This paper introduces a generalized federated prompt-tuning framework for practical scenarios where local datasets are multi-modal and exhibit different distributional patterns of missing features at the input level.~The proposed framework bridges the gap between federated learning and multi-modal prompt-tuning which have traditionally focused on either uni-modal or centralized data.~A key challenge in this setting arises from the lack of semantic alignment between prompt instructions that encode similar distributional patterns of missing data across different clients. To address this, our framework introduces specialized client-tuning and server-aggregation designs that simultaneously optimize, align, and aggregate prompt-tuning instructions across clients and data modalities.~This allows prompt instructions to complement one another and be combined effectively.~Extensive evaluations on diverse multimodal benchmark datasets demonstrate that our work consistently outperforms state-of-the-art (SOTA) baselines.
\vspace{-4mm}

\end{abstract}

\section{Introduction}
\label{sec:intro}

The recent emergence of exceedingly large models~\cite{Bommasani2021FoundationModels} with immense scale and versatility due to being pre-trained on large swaths of generic data has resulted in a paradigm shift in machine learning (ML). Most task-specific models are now fine-tuned versions of those large models instead of being created from de novo learning processes. For example, most state-of-the-art (SOTA) models solving specific NLP tasks are fine-tuned versions of existing large pre-trained or foundation models such as BERT~\cite{devlin-etal-2019-bert}, GPT~\cite{gpt-brown-2020}, and T5~\cite{2020t5}. Likewise, SOTA models for various computer vision tasks have also been derived from foundation models such as the vision transformer~\cite{dosovitskiy2020vit} pre-trained on extensive datasets like ImageNet~\cite{imagenet}. The standard fine-tuning approach, however, assumes centralized data which is not always possible in practical scenarios where fine-tuning data are scattered across multiple private siloed systems. 


To overcome this siloed data challenge, federated fine-tuning has recently been considered to aggregate the fine-tuned updates across private devices running their own (local) fine-tuning processes~\cite{chen2022fedtune,nguyen2023begin,weng2024probabilistic}.~This approach can be further combined with an existing parameter-efficient tuning method called prompt tuning~\cite{lester-etal-2021-power}.~This method focuses on creating lightweight signals such as additional tokens that are prepended to the input of the pre-trained models~\cite{maria2021fewshot,zhou2022coop, zhou2022cocoop,Saito_2023_CVPR}, providing them valuable contextual information to perform specific tasks more effectively.~Local sets of prompts can then be aggregated using the conventional parameter aggregation procedure in FL~\cite{guo2022promptfl,zhao2023fedprompt,Yang_2023_ICCV,lu2023fedclip,weng2024probabilistic}.

However, such federated prompt-tuning techniques are restricted to learning with uni-modal data. In more practical scenarios, advances in sensor and device manufacturing increasingly enable the capture of multiple data modalities (e.g., text, images), leading to richer datasets for fine-tuning~\cite{brunete2021smart}.~Leveraging such diverse multi-modal information can significantly enhance model performance, as demonstrated by the success of multi-modal foundation models like CLIP~\cite{CLIP} and ViLT~\cite{kim2021vilt}.~This has further motivated research on multi-modal federated learning~\cite{chen2022fedmsplit, phung2024mifl,yu2024fedinmm, Nguyen2024FedMACTP} and fine-tuning pre-trained multi-modal LLMs, i.e., multi-modal federated fine-tuning~\cite{zhang2024mllm}.

Despite these advances, multi-modal federated learning research often adopts specific local architectures which cannot be learned via fine-tuning existing pre-trained multi-modal foundation models~\cite{chen2022fedmsplit, phung2024mifl,yu2024fedinmm, Nguyen2024FedMACTP}.~Existing multi-modal federated fine-tuning research on the other hand assumes that all local datasets have the same set of modalities~\cite{zhang2024mllm}.~As such, these approaches are not suitable for situations where different clients might collect data using different types of sensors.~Furthermore, the multi-modal sensor data across clients might even have different missing patterns of data entries.~This leads to local datasets with both intra- and inter-heterogeneities.~Intra-heterogeneity occurs when individual samples within a single dataset have missing entries for different subsets of features.~Inter-heterogeneity arises when different clients have different distributions over their data-missing patterns.

\begin{figure}[bt]
\centering
\includegraphics[width=0.9\linewidth]{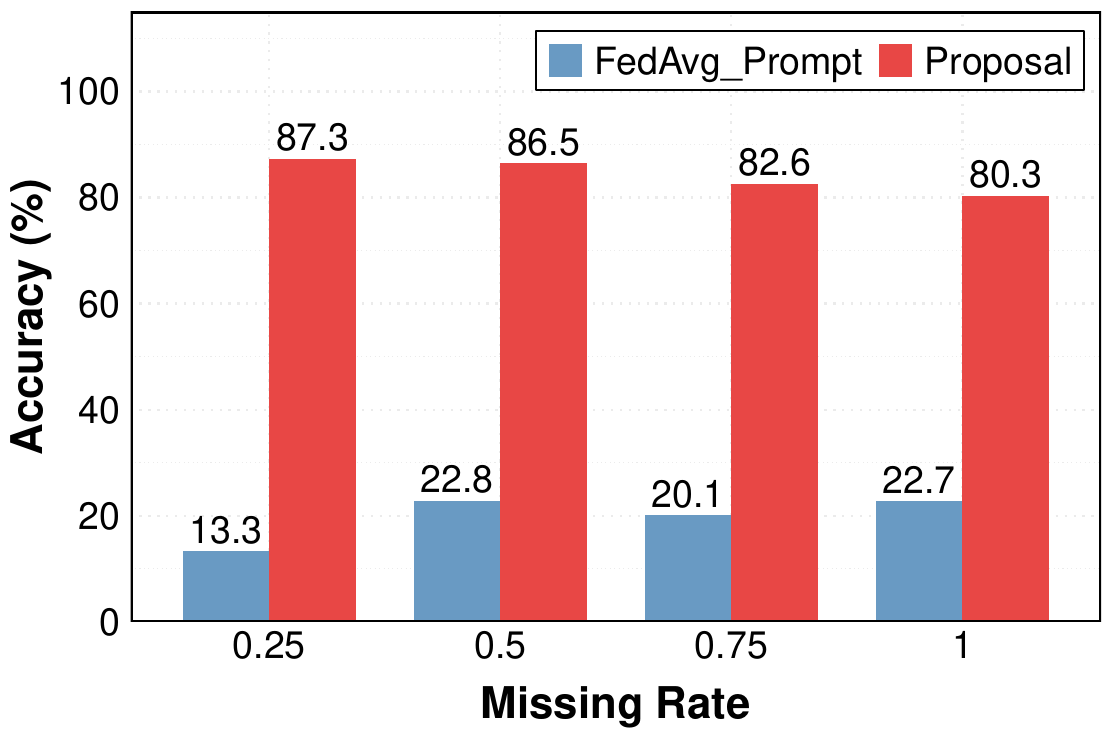}\vspace{-3mm}
\caption{Our approach with prompt alignment outperforms FedAvg prompt-tuning (w/o alignment) across settings with different data missing rates.}
\vspace{-6mm}
\label{fig:motivation}
\end{figure}

There also exists recent work in prompt-tuning that learns a specific prompt for each subset of missing modalities~\cite{lee2023cvpr}.~Input samples having the same set of missing modalities are then augmented with the same (learnable) prompt during training, specializing it to complement for the corresponding missing modalities.~This approach is, however, restricted to centralized data setting (i.e., no inter-heterogeneities) and does not generalize well to federated settings where different sets of local prompts might be biased towards different distributional patterns of missing data.~A naive aggregation of these prompts might therefore collapse them into less informative prompts which are ignorant of the heterogeneities in missing-data patterns across clients, resulting in poor performance (see Fig.~\ref{fig:motivation}).

To bridge this gap between multi-modal prompt-tuning and federated learning, we propose a multimodal federated prompt-tuning approach that optimizes and aggregates two distinct sets of (local) prompts for each client: inter-client and intra-client prompts.~Inter-client prompts breakdown and encode prototypical distribution patterns of missing input data, which can be aligned and aggregated across clients to represent shared distributional information. In contrast, intra-client prompts encode information tailored to support specific local subsets of missing modalities, independent of the distributional patterns of missing data at each client.~This is achieved via the following contributions:\vspace{1mm}



\noindent {\bf 1.}~We develop a local input-adaptive fine-tuning algorithm which identifies and updates a subset of inter- and intra-client prompts that are most relevant to each individual sample.~This is formulated as prompt-retrieval task which focuses on  learning mapping functions that push an input towards its relevant prompts on an embedding space while pulling away from irrelevant prompts on an embedding space. This helps decompose and distill local fine-tuning knowledge into different parts of the prompt set, preventing learned instructions for different fine-tuning patterns from being overloaded on the same prompts (Section 3.2).\vspace{1mm}

\noindent {\bf 2.}~We develop a server-side model for aggregating and aligning inter-client prompts across clients.~Both generation and alignment are framed as part of a clustering task, where inter-client prompts encoding similar input-level distribution patterns of missing data are identified and merged into more comprehensive and versatile prompt instructions. The empirical effectiveness of these combined prompts in reducing local training losses is used as a learning feedback to optimize this clustering model (Section 3.3). \vspace{1mm}

\noindent {\bf 3.}~We evaluate and demonstrate the effectiveness of our proposed approach through extensive experiments on two benchmark datasets, MM-IMDB~\cite{arevalo2017IMDB} and UPMC Food-101~\cite{upmcfood101}, comparing its performance against a variety of existing multimodal federated learning methods as well as repurposed prompt-tuning baselines for federated settings.~The reported results show that our method consistently and significantly outperforms the baselines across various modality-missing scenarios, achieving new SOTA performance for multi-modal federated learning (Section~\ref{sec:exp}).


\noindent A bird-view of our proposed framework, \underline{Fed}erated \underline{Pr}ompt-Tuning with Heterogeneous and \underline{I}ncomplete \underline{M}ultimodal Cli\underline{e}nt Data (\ourmethod), is also provided in Figure~\ref{fig:framework_overview}.
\vspace{-5mm}

\section{Related Works}
\label{sec:related_works}

\noindent \textbf{Multi-Modal Federated Learning.}~Federated learning (FL)~\cite{mcmahan2017fedavg,NghiaNIPS19,NghiaICML19b,NghiaICML19a} is a collaborative learning paradigm that enables multiple devices to train a shared model while keeping local data private.~This design addresses concerns over privacy, security, and inefficiency in transferring large datasets~\cite{zhang2022data, wang2022communication}. While FL has been widely studied~\cite{NghiaAAAI19,li2019convergence, t2020personalized, li2020federated, fallah2020personalized, cho2022heterogeneous, pham2023sem, nguyen2023cadis, nguyen2022feddrl,NghiaUAI23,NghiaUAI23b,sim2023incentives}, most existing methods focus on uni‑modal data (e.g., image or text). However, with the rapid advancement of mobile phones and IoT devices~\cite{brunete2021smart}, local data increasingly comes from multiple modalities. Integrating such heterogeneous signals enables richer, more robust representations~\cite{kaur2021comparative, xiong2022unified}, motivating growing interest in multi‑modal FL. Yet many current approaches rely on custom architectures that cannot exploit pre‑trained multi‑modal models. More recent studies explore federated fine‑tuning of pre‑trained multi‑modal models, but they typically assume each client holds the same set of modalities~\cite{zhang2024mllm}, limiting applicability in realistic heterogeneous environments.

\noindent \textbf{Missing Data in Multi‑Modal FL.} In real‑world settings, local devices collect data from different types of sensors, leading to datasets with varying feature modalities (inter‑heterogeneity). Moreover, sensor malfunctions or data corruption can cause missing features, with devices exhibiting distinct missing patterns (intra‑heterogeneity). Existing solutions either treat each modality independently~\cite{zhang2020deep, zhou2022missing} or are restricted to centralized learning~\cite{yu2020optimal, chen2020hgmf, poklukar2022geometric}, both overlooking inter‑modal dependencies. Recent FL methods address inter‑heterogeneity~\cite{chen2022fedmsplit, phung2024mifl} but not intra‑heterogeneity, while approaches like FedInMM~\cite{yu2024fedinmm} and FedMAC~\cite{Nguyen2024FedMACTP} tackle intra‑client missing data but ignore cross‑client alignment, limiting effectiveness. A related line of work, including FedMSplit~\cite{chen2022fedmsplit} and CACMRN~\cite{xiong2023client}, handles modality‑incomplete data under a personalized FL framework, where each client retains its model. This paradigm differs from standard FL, which aims to improve a single shared model, making these methods not directly comparable.

\noindent \textbf{Addressing Missing Data with~Prompt‑Tuning.} Prompt‑tuning~\cite{maria2021fewshot, zhou2022coop, zhou2022cocoop, Saito_2023_CVPR} provides an effective way to adapt pre‑trained models by learning prompts that add contextual information to input embeddings. Recent work has extended this idea to missing‑modality scenarios by learning a prompt for each subset of missing features~\cite{lee2023cvpr, khattakMaPLe}, so that inputs with similar missing patterns are augmented with suitable prompts. However, these methods are developed for centralized training and do not address the additional heterogeneity introduced in FL, where prompts must be aligned and aggregated across clients. Designing effective alignment and aggregation strategies for such prompts is therefore a key challenge, as demonstrated in our framework.
\vspace{-2mm}

\section{Multi-Modal Federated Prompt-Tuning}
\label{sec:methodology}

\begin{figure*}[tb]
\centering
\includegraphics[width=\linewidth]{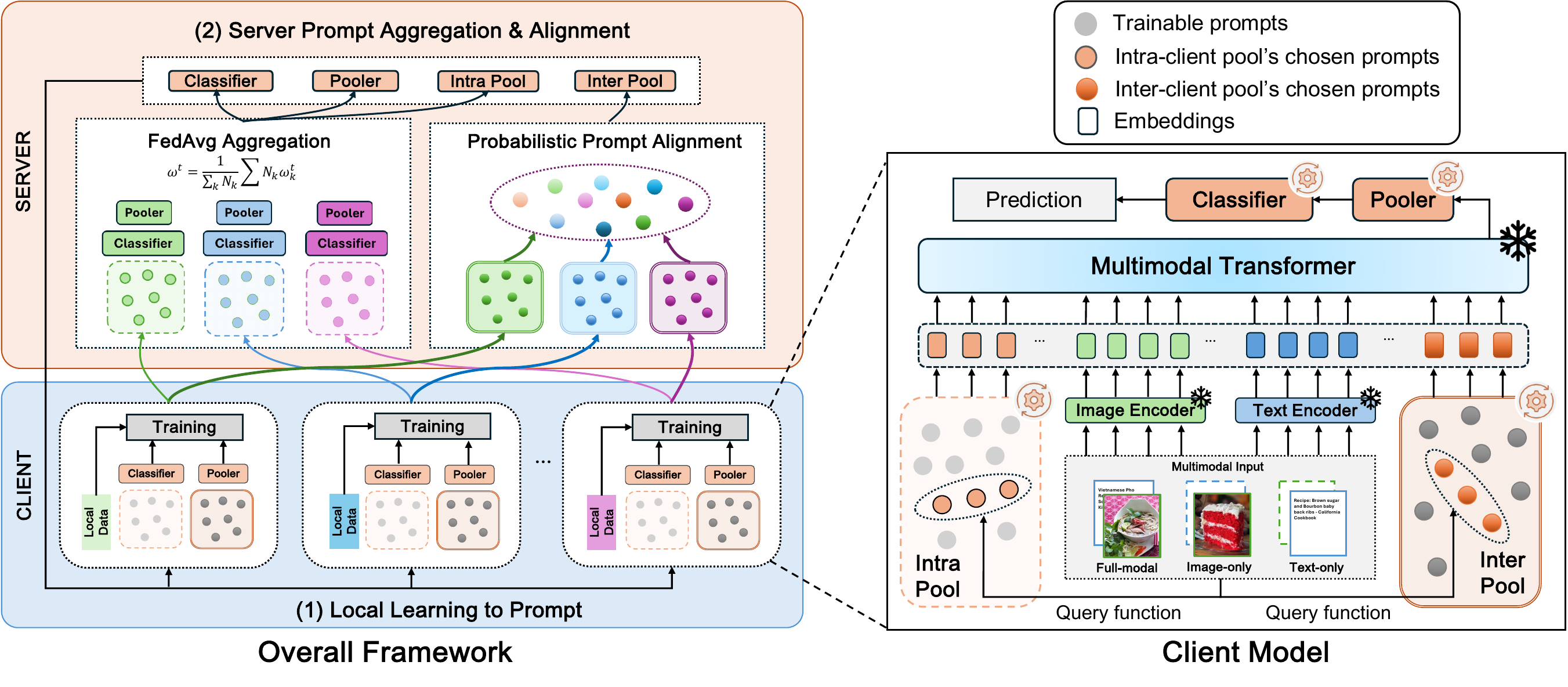}
\vspace{-10pt}
\caption{Overview of the proposed multi-modal federated prompt-tuning framework -- see Alg.~\ref{alg:FPT}. Each client maintains two (learnable) sets of intra- and inter-client prompts. At the beginning of each iteration, each client performs local training via Eq.~\eqref{eq:5} and Eq.~\eqref{eq:6}. Local sets of intra- and inter-client prompts are then sent to the server for aggregation -- see Eq.~\eqref{eq:8}--Eq.~\eqref{eq:12}.}
\label{fig:framework_overview}
\vspace{-12pt}
\end{figure*}

Multimodal federated learning faces significant challenges due to missing data modalities, leading to both inter-client and intra-client heterogeneities that complicate knowledge aggregation and degrade overall model performance -- see Fig.~\ref{fig:motivation}.~To address this challenge (Section~\ref{sec:formulation}), we present a federated prompt-tuning framework based on the pre-trained Vision and Language Transformer (ViLT) model, which comprises a prompt-based client design and a server-side prompt-aggregation algorithm. The client design features a fine-tuning algorithm that learns and decomposes the fine-tuned knowledge into separate sets of inter-client and intra-client prompts specialized for fine-tuning at different input-level patterns of missing data and input-agnostic patterns of missing modalities (Section~\ref{sec:client}).~The server-side algorithm then learns alignment between inter-client prompts specializing for similar input-level patterns of missing data and aggregates them to create more versatile prompts. This is also combined with a standard \textsc{FedAvg}-based aggregation of intra-prompts specializing for input-agnostic patterns of missing modalities (Section~\ref{sec:server}) -- see Fig.~\ref{fig:framework_overview}.
\vspace{-2mm}


\subsection{Problem Formulation}
\label{sec:formulation}
In multimodal federated learning (MMFL), we assume there are $n$ clients and $r$ data modalities (e.g., text, image). Each client $t \in [n]$\footnote{$[n] \triangleq \{1, 2, \ldots, n\}$.} has access to a private dataset $D_t \triangleq \{\boldsymbol{x}(M_{t, s}), z_{t, s}\}_{s=1}^{m}$, where $\boldsymbol{x}(M_{t, s})$ denote the observed features for the set of modalities, $M_{t,s} \subseteq [r]$ recorded for the $s$-th data point of the $t$-th client, and $z_{t, s}$ denotes the corresponding target label. The goal is to learn a multimodal model $\boldsymbol{w}_\ast$ using all the private datasets $\{D_t\}_{t=1}^n$ without centralizing them. That is,
\begin{eqnarray}
\hspace{-5mm}\boldsymbol{w}_\ast \hspace{-2mm}&=&\hspace{-2mm} \underset{\boldsymbol{w}}{\arg\min}\left\{ L(\boldsymbol{w}) \triangleq \frac{1}{n}\sum_{t=1}^n L_t(\boldsymbol{w})\right\} \ \text{with}\ \nonumber\\
L_t(\boldsymbol{w}) \hspace{-2mm}&\triangleq&\hspace{-2mm} \frac{1}{n}\sum_{s=1}^{m} \ell\Big(F\big(\boldsymbol{x}\big(M_{t,s}\big); \boldsymbol{w}\big), z_{t,s}\Big) \label{eq:1}
\end{eqnarray}
where $F(\boldsymbol{x}\big(M_{t,s}\big); \boldsymbol{w})$ denotes the multimodal prediction model with parameter $\boldsymbol{w}$ and $\ell(F(\boldsymbol{x}\big(M_{t,s}\big); \boldsymbol{w}),z_{t,s})$ denotes the individual loss of $\boldsymbol{w}$ on the $s$-th data point.~Previous work~\cite{chen2022fedmsplit, phung2024mifl, yu2024fedinmm, Nguyen2024FedMACTP} in MMFL has largely focused on designing $F(\boldsymbol{x}\big(M_{t,s}\big); \boldsymbol{w})$ without leveraging existing knowledge in pre-trained foundation model.~Instead, we let $F(\boldsymbol{x}\big(M_{t,s}\big); \boldsymbol{w})$ be a prompt-tuned version of a pre-trained transformer-based model,
\begin{eqnarray}
\hspace{-9mm}F\Big(\boldsymbol{x}(M); \boldsymbol{w}\Big) \hspace{-2mm}&\triangleq&\hspace{-2mm} F_c\Big(F_p\Big(F_e\big(\boldsymbol{x}(M)\big)\ \circ\  \boldsymbol{w}_p\Big); \boldsymbol{w}_c\Big)\label{eq:2}
\end{eqnarray}
where $F_c(.; \boldsymbol{w}_c)$ is the prediction head parameterized with $\boldsymbol{w}_c$, $F_e(\boldsymbol{x}(M)) = F_e(\boldsymbol{x}_1) \circ \ldots \circ F_e(\boldsymbol{x}_{|M|})$ where $F_e(\boldsymbol{x}_i)$ denote the tokenized input sequence of each unimodal feature $\boldsymbol{x}_i$ in $\boldsymbol{x}(M) \triangleq (\boldsymbol{x}_1, \boldsymbol{x}_2, \ldots, \boldsymbol{x}_{|M|})$, $F_p(.)$ denotes the frozen multimodal pre-trained model -- e.g., ViLT~\cite{kim2021vilt}  -- and $\boldsymbol{w}_p$ denotes the set of (learnable) prompts. Here, $\circ$ denotes the set concatenation operator and $\boldsymbol{w} = (\boldsymbol{w}_c, \boldsymbol{w}_p)$ denotes the full set of tuning parameters. Due to the decentralized and private nature of the local datasets $\{D_t\}_{t=1}^n$, solving Eq.~\eqref{eq:1} directly is not possible. Instead, following the federated learning~\cite{mcmahan2017fedavg} paradigm, its optimal solution $\boldsymbol{w}_
\ast$ is approximated via a number of iterations over two interleaving client- and server-side steps:
\begin{eqnarray}
\hspace{-14mm}\boldsymbol{w}^{(h+1)}_t &=& \mathrm{update}\Big(L_t, \boldsymbol{w}^{(h)}_\ast\Big)\ \forall t \in [n] \ ,\label{eq:3a}\\
\hspace{-14mm}\boldsymbol{w}^{(h+1)}_\ast &=& \mathrm{aggregate}\Big(\boldsymbol{w}_1^{(h +1)}, \ldots, \boldsymbol{w}_n^{(h +1)}\Big), \label{eq:3b}
\end{eqnarray}
where $\boldsymbol{w}_t^{(h)}$ denote the estimation of the local tuning parameter $\boldsymbol{w}_t$ of client $t$ after $h$ iterations. For example, the $\mathrm{update}$ step can be a vanilla gradient descent (GD) optimization of the prompt $\boldsymbol{\omega}$ and the $\mathrm{aggregate}$ step denotes the average of the locally optimized prompts. Such vanilla approach however does not account for both the inter- and intra-heterogeneities among local prompts (see Section~\ref{sec:intro}) which ultimately risks aggregating incompatible prompts and degrade the overall performance (see Section~\ref{sec:exp}). To avoid this, we will introduce novel designs for client- and server-prompts as detailed in the followings.

\subsection{Client Design}
\label{sec:client}
To enable effective prompt-aggregation across heterogeneous clients, we need to separate the fine-tuning treatment for input-level patterns of missing data and input-agnostic patterns of missing modalities.~This can be achieved via further parameterizing $\boldsymbol{w}_p = (\boldsymbol{w}_p^{inter}, \boldsymbol{w}_p^{intra})$ with two distinct subsets of $\tau$ inter- and intra-client prompts:
\begin{eqnarray}
\hspace{-17.5mm}\boldsymbol{w}^{inter}_p &\triangleq& \Big\{ \boldsymbol{p}^{inter}_1,\ \boldsymbol{p}^{inter}_2,\ \dots,\ \boldsymbol{p}^{inter}_\tau \Big\} \ ,\\
\hspace{-17.5mm}\boldsymbol{w}_p^{intra} &\triangleq& \Big\{ \boldsymbol{p}^{intra}_1,\ \boldsymbol{p}^{intra}_2,\ \dots,\ \boldsymbol{p}^{intra}_\tau \Big\} \ ,
\end{eqnarray}
which are learned separately.~The learned inter-client prompts $\boldsymbol{w}^{inter}_p$ across clients will be sent to the server which clusters them into different groups based on their specialization for similar input-level patterns of missing data -- Section~\ref{sec:server}.~Local sets of intra-client prompts $\boldsymbol{w}^{intra}_p$ can be averaged via \textsc{FedAvg} \cite{mcmahan2017fedavg} since their specialization is for input-agnostic patterns of missing modalities. ~Intuitively, the aggregation mechanisms will determine how different aspects of fine-tuning knowledge are encoded across the prompt sets.~For example, if knowledge regarding different input-level patterns of missing data is (incorrectly) incorporated into $\left\{\boldsymbol{p}^{intra}_i\right\}_{i=1}^\tau$, they will be averaged out and collapsed into less informative prompts due to the non-clustering aggregation mechanism of $\left\{\boldsymbol{p}^{intra}_i\right\}_{i=1}^\tau$. 

As this potential loss of information will decrease the overall performance, the (learnable) prompt parameters will be updated to prevent this from happening.~Otherwise, if (common) knowledge patterns regarding input-agnostic patterns of missing modalities are (incorrectly) incorporated into $\left\{\boldsymbol{p}^{inter}_i\right\}_{i=1}^\tau$, their modeling bandwidth for inter-client heterogeneities will be reduced, which will also hamper the performance.~This will, in turn, generate a prompt update signal to avoid this as well.
Furthermore, within each set of inter- and intra-client prompts, there might also be different patterns that need to be separated and internalized into different subset of prompts.

To enable this, we develop an input-adaptive learning mechanism for both sets of prompts which identifies and updates the most relevant subset of prompts for each (multimodal) input sample.~This is formulated as an information-retrieval task, learning a pair of key and query functions, $k\left( \boldsymbol{p} \right)$ and $q(\boldsymbol{x}(M))$, which transform the prompt $\boldsymbol{p}$ and input $\boldsymbol{x}$ with the set $M$ observed modalities, respectively, into a shared metric space. We can then define a  geometric input-prompt distance, e.g., $d(\boldsymbol{x}(M), \boldsymbol{p}) = \mathrm{cos}(q(\boldsymbol{x}(M)), k\left(\boldsymbol{p}\right))$ which can be used to measure their contextual relevance. To avoid loading diverse fine-tuning knowledge into a single prompt, we regularize the (local) loss function to penalize it when a less relevant prompt is selected as fine-tuning instruction for an input $\boldsymbol{x}(M)$: 
\begin{eqnarray}
\boldsymbol{w}_t 
&\triangleq& \underset{\boldsymbol{w}}{\arg\min}\ \  L'_t(\boldsymbol{w}) \quad \text{where}\quad\label{eq:5}\\
L'_t\Big(\boldsymbol{w}\Big) &\triangleq& \sum_{s=1}^{m}\Bigg\{ \ell\Big(F\Big(\boldsymbol{x}\big(M_{t,s}\big); \boldsymbol{w}'\Big), z_{t,s}\Big)\Bigg\}\nonumber \\
&+& \sum_{s=1}^{m}\Bigg\{r\Big(\boldsymbol{x}(M_{t,s}), \boldsymbol{w}'_p\Big)\Bigg\}  \ , \label{eq:6}
\end{eqnarray}
where $\ell(\boldsymbol{w}')$ is defined in Eq.~\eqref{eq:1} and $\boldsymbol{w}' = (\boldsymbol{w}_c, \boldsymbol{w}'_p)$ denote the customized set of tuning instructions for each input $\boldsymbol{x}(M_{t,s})$ where $\boldsymbol{w}'_p= (\boldsymbol{w}_p^{inter'}, \boldsymbol{w}_p^{intra'})$ such that $\boldsymbol{w}_p^{inter'} \subseteq \boldsymbol{w}_p^{inter}$ and $\boldsymbol{w}_p^{intra'} \subseteq \boldsymbol{w}_p^{intra}$ are the sets of $\kappa$ most relevant prompts to $\boldsymbol{x}(M_{t,s})$ in  $\boldsymbol{w}_p^{inter}$ and $\boldsymbol{w}_p^{intra}$.~The regularizer $r(\boldsymbol{x}(M), \boldsymbol{w}'_p)$ is then defined as
\begin{eqnarray}
r\Big(\boldsymbol{x}\big(M\big), \boldsymbol{w}'_p\Big) &\triangleq& \sum_{\boldsymbol{p} \in \boldsymbol{w}^{'}_p}\Bigg\{ d\Big(\boldsymbol{x}\big(M\big),\ \boldsymbol{p} \Big)\Bigg\} \ .\label{eq:7}
\end{eqnarray}
Intuitively, the regularizer penalizes the total distance between the (multimodal) input $\boldsymbol{x}(M)$ and the $\kappa$ closest prompts according to the aforementioned distance metric $d(\boldsymbol{x}(M),\boldsymbol{p})$. Minimizing both the original local loss and this regularizer allows the local model to simultaneously learn to (1) internalize fine-tuning patterns of inter- and intra-heterogeneities to minimize the original loss $L_t(\boldsymbol{w})$; and (2) ensure that selected prompts for a given input will be close to it to reduce the penalty $r(\boldsymbol{x}(M), \boldsymbol{w}'_p)$. A pseudocode of this client design is detailed in Suppl.~\ref{app:a}.





\begin{figure}[tb]
\centering
\includegraphics[width=0.95\linewidth]{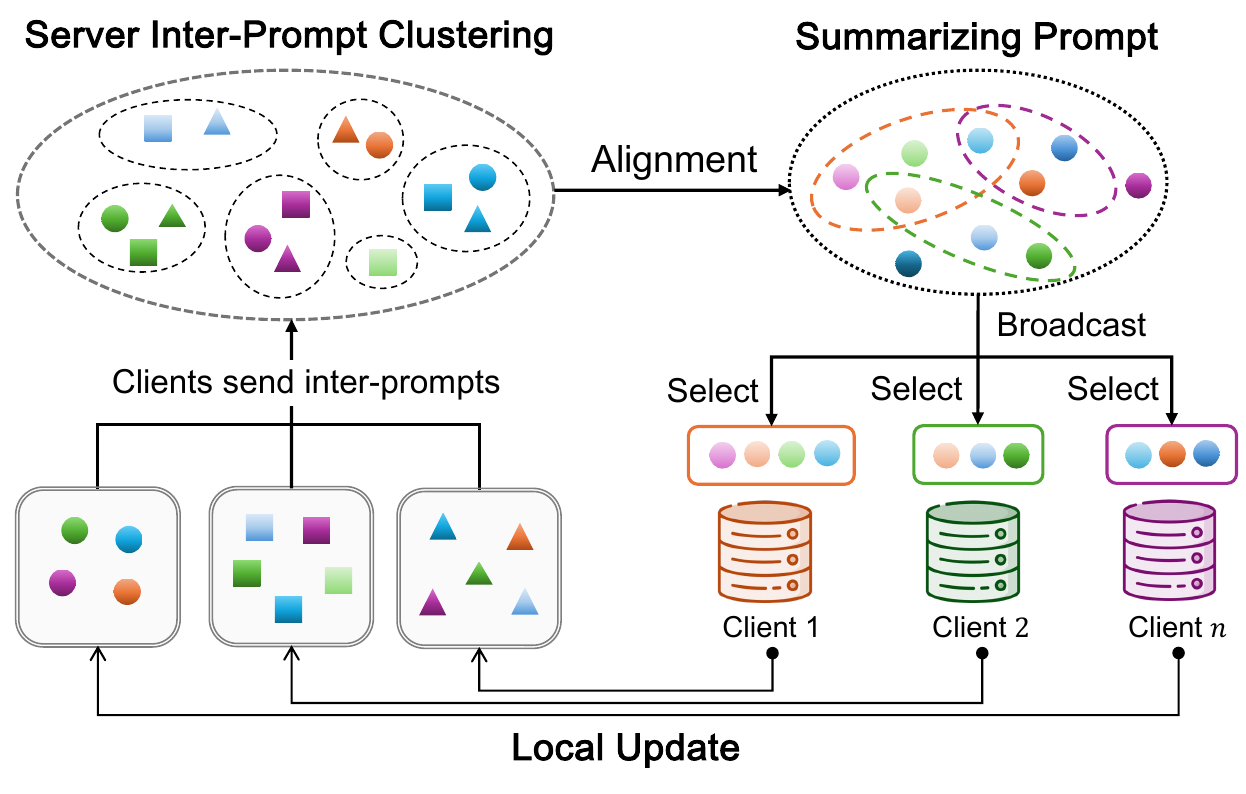}
\vspace{-5pt}
\caption{Workflow of the prompt-alignment algorithm. At each iteration, each client samples a subset of summarizing prompts using its query and key functions. The client then performs a local update, resulting in heterogeneous prompt sets, which are subsequently sent to the server to be clustered and summarized into new summarizing prompt sets for the next iteration.}
\label{fig:prompt-alignment}
\end{figure}

\subsection{Server Aggregation}
\label{sec:server}
We will now establish a principled algorithm to align and aggregate inter-client prompts specializing in similar input-level patterns of missing data\footnote{This section only describes the aggregation algorithm for inter-client prompts since intra-client prompts are specialized for input-agnostic patterns of missing modalities and  can be aggregated via  \textsc{FedAvg}~\cite{mcmahan2017fedavg}.}. This is not straight-forward since inter-client prompts at the same position in the prompt sets across different clients might not specialize for similar patterns of missing data due to client heterogeneities: certain missing-data patterns are not observed at certain clients.  


To find the correct alignment, we formulate it as a clustering optimization task. Suppose each client $t$ communicates a set of inter-client prompts $\boldsymbol{w}_t^{inter} = (\boldsymbol{p}_t^i)_{i=1}^\tau$. Each potential clustering of these prompts thus defines a candidate alignment, i.e., prompts in the same cluster are considered aligned and can be aggregated.~A (learnable) cluster center can then be associated with the aggregated prompt -- see Fig.~\ref{fig:prompt-alignment}.~Both the cluster assignment and centers can be found via solving the following optimization task with discrete alignment variables $\alpha_t^{pq} \in \{0, 1\}$ and continuous center variables $\boldsymbol{\theta}_q$ for each cluster $q \in [n\times\tau]$:
\begin{eqnarray}
\hspace{-7mm}\underset{\boldsymbol{\alpha}, \boldsymbol{\theta}, \gamma}{\min}\ \  G\Big(\boldsymbol{\alpha}, \boldsymbol{\theta}, \gamma\Big) \hspace{-3mm}&\triangleq&\hspace{-3mm} \sum_{t=1}^n\sum_{p=1}^\tau\sum_{q=1}^{n \times \tau} \alpha_{t}^{pq}\cdot c\Big(\boldsymbol{p}_t^p, \boldsymbol{\theta}_q, ; \gamma\Big) \label{eq:8}\\
\hspace{-3mm}&\mathrm{s.t.}&\hspace{-3mm}  \alpha_t^{p, 1} + \ldots + \alpha_t^{p, n\times\tau} \ = \ 1 \ ,\label{eq:9}
\end{eqnarray}
where $\boldsymbol{\alpha} \triangleq (\alpha_t^{p, q})$ and $\alpha_t^{p,q} \in \{0,1\} \ \forall(t, p, q)$ indicates whether the $p$-th inter-client prompt of client $t$ is matched to the $q$-th cluster. The corresponding cluster center is then optimized to have minimum total distance to all cluster members according to a (learnable) cost function $c(\boldsymbol{p}_t^p, \boldsymbol{\theta}_q)$. The optimized cluster center can then be used as the aggregated or summarizing prompt.~Minimizing Eq.~\eqref{eq:8} thus allows us to align and summarize/aggregate the inter-client prompts across clients simultaneously.~Note that Eq.~\eqref{eq:9} enforces a constraint that intuitively indicates two prompts from the same clients should not be matched to the same cluster and get aggregated.~This is because local prompts from the same client have already been learned to encode different fine-tuning aspects.~Aggregating them will therefore collapse them into less informative ones.

In addition, some summarizing/aggregated prompts might be more generic than others and can be used to provide fine-tuning instructions to a wider range of (multimodal) inputs. The update of such prompts should be prioritized during training and to achieve that, we need a mechanism to capture this \emph{prompt popularity}. To achieve this, we augment the optimization task in Eq.~\eqref{eq:8} above with a regularizer that penalizes alignment to \emph{less popular} prompts,
\begin{eqnarray}
\hspace{-2mm}R\Big(\boldsymbol{\alpha}, \zeta\Big) \hspace{-2mm}&\triangleq&\hspace{-2mm} \sum_{t=1}^n\sum_{p=1}^\tau\sum_{q=1}^{n\times\tau} \Big(\alpha_t^{p, q} \cdot \log U(\boldsymbol{\theta}_q; \zeta)\Big) \\ 
\hspace{-2mm}&+&\hspace{-2mm} \sum_{t=1}^n\sum_{p=1}^\tau\sum_{q=1}^{n\times\tau}\Big(1 - \alpha_t^{p, q}\Big) \hspace{-1mm}\cdot\hspace{-1mm} \log\Big(1 - U(\boldsymbol{\theta}_q; \zeta)\Big), \nonumber\label{eq:10}
\end{eqnarray}
where $U(\boldsymbol{\theta}_q; \zeta)$ is the popularity function of $\boldsymbol{\theta}_q$, which is parameterized by learnable parameter $\zeta$.~For example, $U(\boldsymbol{\theta}_q; \zeta) = \sigma(g(\boldsymbol{\theta}_q; \zeta)) \in (0,1)$\footnote{$\sigma(.)$ denotes the sigmoid function.} where $g(\boldsymbol{\theta}_q; \zeta)$ is a deep neural net parameterized with $\zeta$. Putting Eq.~\eqref{eq:10} and Eq.~\eqref{eq:8} together, we arrive at our final loss function,
\begin{eqnarray}
\underset{\boldsymbol{\alpha}, \boldsymbol{\theta}, \gamma, \zeta}{\mathrm{minimize}}\ \  \Big\{G\Big(\boldsymbol{\alpha}, \boldsymbol{\theta}, \gamma\Big) \ +\  R\Big(\boldsymbol{\alpha}, \zeta\Big)\Big\} \ , \label{eq:11}
\end{eqnarray}
which can be optimized via alternating between (1) optimizing $( \boldsymbol{\theta}, \zeta, \gamma)$ while fixing $\boldsymbol{\alpha}$; and (2) optimizing $\boldsymbol{\alpha}$ while fixing  $(\boldsymbol{\theta}, \zeta, \gamma)$. The first optimization sub-task is straight-forward while the latter is less trivial due to the constraint as well as the discrete nature of the optimizing variables. To sidestep the technical challenge in solving for $\boldsymbol{\alpha}$ while fixing  $(\boldsymbol{\theta}, \zeta, \gamma)$, we approximate it by optimizing for $\boldsymbol{\alpha}_t = \{\alpha_t^{p, q}\}_{p, q}$ while also fixing $\boldsymbol{\alpha}_{-t} = \{\alpha_{-t}^{p, q}\}_{p, q}$. This results in a minimum weighted linear sum task,
\begin{eqnarray}
\underset{\boldsymbol{\alpha}_t}{\min} \ \  L(\boldsymbol{\alpha}_t) \hspace{-2mm}&\triangleq&\hspace{-2mm} \sum_{p=1}^\tau\sum_{q=1}^{n\times\tau} \Big(\alpha_t^{p,q} \cdot v(p, q)\Big) \ \  \text{where}\label{eq:12}\\
v(p, q) \hspace{-2mm}&\triangleq&\hspace{-2mm} \left(\log\Bigg\{\frac{U\Big(\boldsymbol{\theta}_q; \zeta\Big)}{1 - U\Big(\boldsymbol{\theta}_q; \zeta\Big)}\Bigg\} \ +\  c\Big(\boldsymbol{p}_t^p, \boldsymbol{\theta}_q; \gamma\Big)\right)\nonumber,
\end{eqnarray}
which can be solved effectively in polynomial time using the Hungarian algorithm~\cite{Kuhn55}. An overview and  pseudocode of the server-side aggregation algorithm are presented in Fig.~\ref{fig:prompt-alignment} and Suppl.~\ref{app:a}. A thorough evaluation of this framework is detailed next in Section~\ref{sec:exp}.



\section{Empirical Evaluation}
\label{sec:exp}
This section provides a detailed empirical evaluation of our proposed framework,
\ourmethod. \footnote{Our implementation is available at \url{https://github.com/hangpt01/FedPrime}.}
Our evaluation is conducted on two datasets UPMC Food-101~\cite{upmcfood101} and MM-IMDB~\cite{arevalo2017IMDB} and several baselines adapting existing works in federated fine-tuning (for unimodal data) and multi-modal prompt-tuning (for centralized data) to our novel setting on decentralized multi-modal data with missing modalities and features. Detailed description of the datasets and baselines are provided in Section~\ref{sec:setting}. Our main results and ablation studies are reported in Section~\ref{sec:result}.

\begin{table*}[t]
\centering
\caption{Performance achieved by the baselines on the Food-101 (left) and MM-IMDB (right) datasets under various missing-data settings in the train and test sets. This includes all combinations of the $3$ train scenarios and $5$ test scenarios (i.e., $15$ experiments) per dataset, as previously stated in Section~\ref{sec:setting}. The best and second-best results are highlighted in \textbf{\textcolor{red}{bold red}}, and \textcolor{blue}{blue}, respectively. \textcolor{blue}{Note that in the {\bf Miss Both} training scenarios, \textbf{Test ($\sim$Train)} and \textbf{Test (Miss Both)} represent the same test scenario according to the definition of {\bf $\sim$Train}. The reported result for \textbf{Test (Miss Both)} in this case is the same as that for \textbf{Test ($\sim$Train)}, as indicated by a dash (-).}}
\label{tab:result-food101}
\resizebox{1\linewidth}{!}{%
\begin{tabular}{@{}c|l|rrrrrrrrrr@{}}
\toprule
\multicolumn{2}{c|}{\textbf{Datasets}} &  \multicolumn{5}{c||}{\textbf{ UPMC Food-101}} & \multicolumn{5}{c}{\textbf{ MM-IMDB}} \\ \cmidrule(l){1-12} 
{\multirow{2}{*}{\textbf{Train}}} &
  {\multirow{2}{*}{\textbf{Method}}} &
  \multicolumn{1}{c}{\textbf{\begin{tabular}[c]{@{}c@{}}Test\\ ($\sim$ Train)\end{tabular}}} &
  \multicolumn{1}{c}{\textbf{\begin{tabular}[c]{@{}c@{}}Test\\ (Miss Both)\end{tabular}}} &
  \multicolumn{1}{c}{\textbf{\begin{tabular}[c]{@{}c@{}}Test\\ (Full Modal)\end{tabular}}} &
  \multicolumn{1}{c}{\textbf{\begin{tabular}[c]{@{}c@{}}Test\\ (Text only)\end{tabular}}} &
  \multicolumn{1}{c||}{\textbf{\begin{tabular}[c]{@{}c@{}}Test\\ (Image only)\end{tabular}}} &
  \multicolumn{1}{c}{\textbf{\begin{tabular}[c]{@{}c@{}}Test\\ ($\sim$ Train)\end{tabular}}} &
  \multicolumn{1}{c}{\textbf{\begin{tabular}[c]{@{}c@{}}Test\\ (Miss Both)\end{tabular}}} &
  \multicolumn{1}{c}{\textbf{\begin{tabular}[c]{@{}c@{}}Test\\ (Full Modal)\end{tabular}}} &
  \multicolumn{1}{c}{\textbf{\begin{tabular}[c]{@{}c@{}}Test\\ (Text only)\end{tabular}}} &
  \multicolumn{1}{c}{\textbf{\begin{tabular}[c]{@{}c@{}}Test\\ (Image only)\end{tabular}}} \\ \midrule
  \multicolumn{1}{c|}{\multirow{5}{*}{\begin{tabular}[c]{@{}c@{}}Miss\\ Text\end{tabular}}} &
  \multicolumn{1}{l|}{\texttt{FEDAVG-P}} &
  15.71 ± 2.27 &
  14.90 ± 1.57 &
  21.56 ± 7.81 &
  16.91 ± 0.69 &
  \multicolumn{1}{r||} {15.36 ± 0.43} & 
  \blue{22.42 ± 2.27} &
  \blue{21.89 ± 1.34} &
  \blue{30.78 ± 1.65} &
  \blue{18.40 ± 0.06} &
  14.53 ± 4.56 \\
\multicolumn{1}{c|}{} &
  \multicolumn{1}{l|}{\texttt{FEDMSPLIT-P}} &
  15.62 ± 1.51 &
  17.50 ± 1.97 &
  25.27 ± 8.22 &
  18.78 ± 0.64 &
  \multicolumn{1}{r||} {17.50 ± 1.97} & 
  {21.02 ± 1.89} &
  {19.97 ± 0.74} &
  {24.39 ± 6.08} &
  {14.38 ± 2.76} &
  \blue{18.09 ± 6.13} \\
\multicolumn{1}{c|}{} &
  \multicolumn{1}{l|}{\texttt{FED-INTER}} &
  54.82 ± 19.01 &
  \blue{48.87 ± 24.64} &
  59.17 ± 27.06 &
  \blue{35.13 ± 26.78} &
  \multicolumn{1}{r||}{56.59 ± 15.12} &
  18.25 ± 3.50 &
  16.95 ± 3.57 &
  18.67 ± 7.63 &
  15.03 ± 4.66 &
  18.01 ± 1.95 \\
\multicolumn{1}{c|}{} &
  \multicolumn{1}{l|}{\texttt{FED-INTRA}} &
  \blue{61.71 ± 17.22} &
  48.09 ± 19.12 &
  \blue{62.06 ± 26.98} &
  22.51 ± 5.92 &
  \multicolumn{1}{r||}{\blue{62.64 ± 11.83}} &
  13.38 ± 1.73 &
  12.77 ± 0.85 &
  12.55 ± 1.67 &
  11.31 ± 0.38 &
  14.33 ± 1.80 \\
\multicolumn{1}{c|}{} &
  \multicolumn{1}{l|}{\textbf{\texttt{FED-PRIME}}} &
  \boldred{78.88 ± 0.90} &
  \boldred{80.38 ± 0.65} &
  \boldred{92.12 ± 0.40} &
  \boldred{73.01 ± 4.25} &
  \multicolumn{1}{r||}{\boldred{76.83 ± 1.22}} &
  \boldred{31.92 ± 0.20} &
  \boldred{31.48 ± 0.30} &
  \boldred{37.67 ± 0.04} &
  \boldred{30.60 ± 1.41} &
  \boldred{30.69 ± 1.41} \\ \cmidrule(l){2-12} 
\multicolumn{1}{c|}{} &
  \multicolumn{1}{l|}{\textbf{Improv. (\%)}} &
  \textbf{27.82 $\uparrow$} &
  \textbf{64.48 $\uparrow$} &
  \textbf{48.44 $\uparrow$} &
  \textbf{107.83 $\uparrow$} &
  \multicolumn{1}{r||}{\textbf{22.65 $\uparrow$}} &
  \textbf{42.37 $\uparrow$} &
  \textbf{43.81 $\uparrow$} &
  \textbf{22.35 $\uparrow$} &
  \textbf{66.30 $\uparrow$} &
  \textbf{69.65 $\uparrow$} \\ \midrule
\multicolumn{1}{c|}{\multirow{5}{*}{\begin{tabular}[c]{@{}c@{}}Miss\\ Image\end{tabular}}} &
  \multicolumn{1}{l|}{\texttt{FEDAVG-P}} &
  17.35 ± 4.77 &
  15.12 ± 1.48 &
  16.84 ± 2.37 &
  18.12 ± 6.49 &
  \multicolumn{1}{r||}{14.81 ± 0.24} &
  \blue{27.69 ± 5.97} &
  \blue{22.55 ± 3.06} &
  \blue{31.94 ± 0.98} &
  \blue{23.76 ± 11.72} &
  12.29 ± 0.47 \\
\multicolumn{1}{c|}{} &
  \multicolumn{1}{l|}{\texttt{FEDMSPLIT-P}} &
  74.16 ± 10.56 &
  48.88 ± 10.26 &
  45.64 ± 32.43 &
  \blue{88.65 ± 2.17} &
  \multicolumn{1}{r||} {14.81 ± 0.90} & 
  {19.11 ± 11.33} &
  {16.61 ± 7.22} &
  {18.19 ± 12.55} &
  {18.12 ± 12.30} &
  12.81 ± 1.25 \\
\multicolumn{1}{c|}{} &
  \multicolumn{1}{l|}{\texttt{FED-INTER}} &
  \blue{77.96 ± 11.62} &
  \blue{64.62 ± 10.22} &
  \blue{82.08 ± 7.75} &
  {77.69 ± 12.35} &
  \multicolumn{1}{r||}{\blue{37.56 ± 6.49}} &
  18.79 ± 5.23 &
  17.93 ± 3.60 &
  20.56 ± 3.56 &
  17.67 ± 6.79 &
  \blue{15.47 ± 2.51} \\
\multicolumn{1}{c|}{} &
  \multicolumn{1}{l|}{\texttt{FED-INTRA}} &
  22.84 ± 3.52 &
  20.13 ± 1.72 &
  23.48 ± 1.86 &
  24.46 ± 2.99 &
  \multicolumn{1}{r||}{16.66 ± 1.32} &
  15.75 ± 4.34 &
  14.06 ± 3.16 &
  15.68 ± 4.77 &
  14.53 ± 3.65 &
  11.71 ± 0.42 \\
\multicolumn{1}{c|}{} &
  \multicolumn{1}{l|}{\textbf{\texttt{FED-PRIME}}} &
  \boldred{90.55 ± 0.22} &
  \boldred{79.12 ± 0.49} &
  \boldred{92.89 ± 0.21} &
  \boldred{90.18 ± 0.29} &
  \multicolumn{1}{r||}{\boldred{54.14 ± 2.50}} &
  \boldred{36.08 ± 0.35} &
  \boldred{31.35 ± 0.61} &
  \boldred{38.49 ± 0.56} &
  \boldred{36.91 ± 0.59} &
  \boldred{18.15 ± 0.66} \\ \cmidrule(l){2-12} 
\multicolumn{1}{c|}{} &
  \multicolumn{1}{l|}{\textbf{Improv. (\%)}} &
  \textbf{16.15 $\uparrow$} &
  \textbf{22.44 $\uparrow$} &
  \textbf{13.17 $\uparrow$} &
  \textbf{1.73 $\uparrow$} &
  \multicolumn{1}{r||}{\textbf{44.14 $\uparrow$}} &
  \textbf{30.30 $\uparrow$} &
  \textbf{39.02 $\uparrow$} &
  \textbf{20.51 $\uparrow$} &
  \textbf{55.35 $\uparrow$} &
  \textbf{17.32 $\uparrow$}\\ \midrule
\multicolumn{1}{c|}{\multirow{5}{*}{\begin{tabular}[c]{@{}c@{}}Miss\\ Both\end{tabular}}} &
  \multicolumn{1}{l|}{\texttt{FEDAVG-P}} &
  14.57 ± 1.50 &
  - &
  17.17 ± 4.37 &
  16.40 ± 4.05 &
  \multicolumn{1}{r||}{13.24 ± 0.32} &
  26.45 ± 2.63 &
  - &
  \blue{33.03 ± 2.56} &
  24.12 ± 11.30 &
  20.21 ± 1.98 \\
\multicolumn{1}{c|}{} &
  \multicolumn{1}{l|}{\texttt{FEDMSPLIT-P}} &
  49.15 ± 24.76 &
  - &
  64.78 ± 36.62 &
  \blue{64.62 ± 36.51} &
  \multicolumn{1}{r||} {21.49 ± 7.19} & 
  {24.25 ± 5.02} &
  {-} &
  {26.05 ± 11.17} &
  {26.02 ± 9.64} &
  19.79 ± 6.20 \\
\multicolumn{1}{c|}{} &
  \multicolumn{1}{l|}{\texttt{FED-INTER}} &
  \blue{56.32 ± 21.77} &
  - &
  \blue{69.57 ± 19.41} &
  {45.15 ± 34.09} &
  \multicolumn{1}{r||}{\blue{59.30 ± 10.84}} &
  \blue{26.53 ± 0.90} &
  - &
  31.97 ± 2.22 &
  \blue{29.69 ± 2.21} &
  \blue{21.63 ± 0.77} \\
\multicolumn{1}{c|}{} &
  \multicolumn{1}{l|}{\texttt{FED-INTRA}} &
  49.28 ± 32.87 &
  - &
  56.70 ± 37.90 &
  43.24 ± 34.19 &
  \multicolumn{1}{r||}{49.85 ± 25.44} &
  11.90 ± 0.37 &
  - &
  12.47 ± 0.45 &
  11.46 ± 0.33 &
  12.83 ± 0.92 \\
\multicolumn{1}{c|}{} &
  \multicolumn{1}{l|}{\textbf{\texttt{FED-PRIME}}} &
  \boldred{84.44 ± 2.65} &
  - &
  \boldred{93.64 ± 0.58} &
  \boldred{87.95 ± 0.91} &
  \multicolumn{1}{r||}{\boldred{72.41 ± 3.88}} &
  \boldred{32.01 ± 2.51} &
  - &
  \boldred{38.68 ± 0.65} &
  \boldred{31.00 ± 2.97} &
  \boldred{26.01 ± 0.12} \\ \cmidrule(l){2-12} 
\multicolumn{1}{c|}{} &
  \multicolumn{1}{l|}{\textbf{Improv. (\%)}} &
  \textbf{49.93 $\uparrow$} &
  - &
  \textbf{34.60 $\uparrow$}&
  \textbf{36.10 $\uparrow$} &
  \multicolumn{1}{r||}{\textbf{22.11} $\uparrow$} &
  \textbf{20.66 $\uparrow$} &
  - &
  \textbf{17.11 $\uparrow$} &
  \textbf{4.41 $\uparrow$} &
  \textbf{20.25 $\uparrow$} \\ \bottomrule
\end{tabular}%
}
\begin{tablenotes}
\small
\item \textbf{(*)} \textbf{Improv.} shows the relative performance improvement between our proposal and the second-best. (in percentage).
\end{tablenotes}
\end{table*}

\subsection{Experiment Settings}
\label{sec:setting}

\noindent \textbf{A. Datasets.}~Following the experiment protocol in~\cite{lee2023cvpr}, we evaluate \ourmethod~on two multimodal datasets:\vspace{1mm}

\noindent {\bf 1.}~UPMC Food-101~\cite{upmcfood101} is an image-text classification dataset consisting of noisy image-text pairs collected from Google Image Search. Its classes are aligned with those in the larger, publicly available ETHZ Food-101 dataset~\cite{ethfood101}. For the federated setting, we select only eight most frequent classes from the original UPMC Food-101.

\noindent {\bf 2.}~MM-IMDB~\cite{arevalo2017IMDB} is a movie genre classification dataset with both image and text modalities. Since each movie can belong to multiple genres, the task is originally a multi-label classification problem, where genres are predicted using image, text, or both. For the federated setting, we use only single-label instances and select eight most frequent classes only from the original MM-IMDB.
\vspace{1mm}

\noindent Each dataset is first partitioned into train and test sets with the 80:20 ratio. The (heterogeneous) missing patterns on the train/test data are simulated using the following protocols.\vspace{1mm}

\noindent \textbf{B. Training Data Simulation.}~Following \cite{lee2023cvpr}, we simulate missing data phenomenon within a dataset using a missing rate $\eta\in(0,1)$ as detailed in $3$ simulation cases below:\vspace{1mm} 

\noindent {\bf 1.~Missing Image.}~We drop the image features in $\eta$ percent of the data and keep the remaining $(1-\eta)$ percent complete.\vspace{1mm}

\noindent {\bf 2.~Missing Text.}~We drop the text features in $\eta$ percent of the data and keep the remaining $(1-\eta)$ percent complete.\vspace{1mm}

\noindent {\bf 3.~Missing Both.}~We drop the image component in $\eta/2$ percent of the data, and the text component in the other $\eta/2$ percent, leaving the remaining $(1-\eta)$ percent complete. \vspace{1mm} 

\noindent The missing-simulated train set is then partitioned into $n$ subsets for $n$ clients via random sampling. The resulting client datasets will therefore have heterogeneous distributions over missing-data patterns.\vspace{1mm}


\noindent \textbf{C. Test Data Simulation.} There are $5$ test scenarios, which include: {\bf (1) Full Modal} (no missing data); {\bf (2) Image Only} (drop all text features); {\bf (3) Text Only} (drop all image features); {\bf (4) Miss Both} (run {\bf Missing Both} on test data); and {\bf (5) $\sim$Train} (same simulation as the training data).\vspace{1mm}

\noindent \textbf{D. Metrics.} Following prior protocol in~\cite{lee2023cvpr}, we use F1-macro and classification accuracy to evaluate the following baseline methods on MM-IMDB and UPMC Food-101.\vspace{1mm}

\noindent \textbf{E. Baseline Methods.}~Following~\cite{lee2023cvpr}, we adopt the pre-trained multimodal transformer ViLT~\cite{kim2021vilt} as the backbone of our client design.~Each client model is a fine-tuned version of this backbone which is generated based on its own local dataset.~As no prior work has addressed modality missing in multimodal fine-tuning within federated learning settings, our comparison baselines mostly feature adapted versions of existing work in federated learning and multi-modal prompt-tuning.~This includes adaptations of the latest missing prompt-tuning SOTA method in~\cite{lee2023cvpr} into the context of the classical FedAvg method as well as that of a more recent FL approach, FedMSplit~\cite{chen2022fedmsplit}. These are referred to as \texttt{FEDAVG-P} and \texttt{FEDMSPLIT-P}, respectively. 
In addition, to underscore the importance of separating fine-tuning knowledge across intra- and inter-client prompt sets, we also include two simplified variants of \texttt{FED-PRIME} utilizing either inter-client prompt pooling or intra-client prompt pooling, along with their respective aggregations. We refer to these variants as \texttt{FED-INTER} and \texttt{FED-INTRA}, respectively (see more details in Supplementary~\ref{app:b}).

\subsection{Experimental Results}
\label{sec:result}
This section provide extensive performance comparison between \ourmethod~ and other baseline methods across various settings of missing data in the train and test sets, as previously stated in Section~\ref{sec:setting}. In addition, we also provide ablation studies on (1) performance robustness against various missing rate $\eta$; (2) the ablated impact of having inter- and intra-client prompt sets; and (3) training convergence. 

\subsubsection{Main Results}
\label{sec:res_main}
We compare the performance of \ourmethod~against the other baselines on an extensive set of experiments across all datasets.~This includes all combinations of the $3$ missing simulation scenarios on train data and $5$ test scenarios (with both missing and no missing data) as previously detailed in parts {\bf B} and {\bf C} of Section~\ref{sec:setting}. All results are reported in Table~\ref{tab:result-food101} which show that \ourmethod~consistently outperforms all baselines in all train/test scenarios with various missing-data patterns. The performance on the UPMC Food-101 and MM-IMDB datasets is measured in terms of classification accuracy and F1-macro, respectively. The F1-macro performance on MM-IMDB is relatively low across all settings due to the class imbalance nature of the dataset. In each experiment, the relative performance improvement of \ourmethod~over the second best baseline is reported, which ranges between $1.73\%$ and $107.83\%$ for the UPMC Food-101 dataset and between $4.41\%$ to $69.65\%$ for the MM-IMDB dataset. 
Notably, \texttt{FED-PRIME} significantly outperforms both of its simplified variants \texttt{FED-INTRA} and \texttt{FED-INTER}, which either handle intra- or inter-heterogeneities, in all settings. This highlights the remarkable synergies and essential of handling both types of heterogeneities in multi-modal federated learning with missing data. In addition, it can also be observed that in more than half of the experiment settings, the second best baseline is either \texttt{FED-INTRA} or \texttt{FED-INTER} which significantly outperform both \texttt{FEDAVG-P} and \texttt{FEDMSPLIT-P} in those cases. These observations complement our findings of the isolated impact of these components in our ablation studies in Section~\ref{sec:res_ablated}. 

\begin{figure*}[t]
\centering
\begin{minipage}[t]{0.28\linewidth}
\centering
\includegraphics[height=3.8cm]{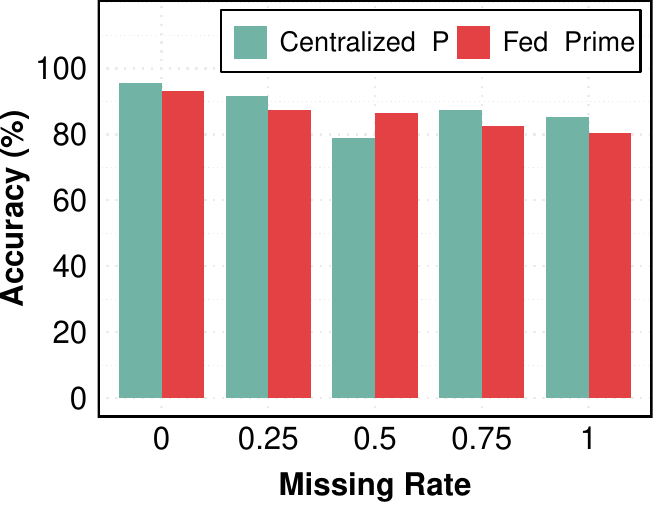}
\caption{~Performance comparison on UMPC Food-101 under {\bf Miss Both} scenarios with various missing rates.}\vspace{-2mm}
\label{fig:miss_both_mean}
\end{minipage}
\hfill
\begin{minipage}[t]{0.3\linewidth}
\centering
\includegraphics[height=3.8cm]{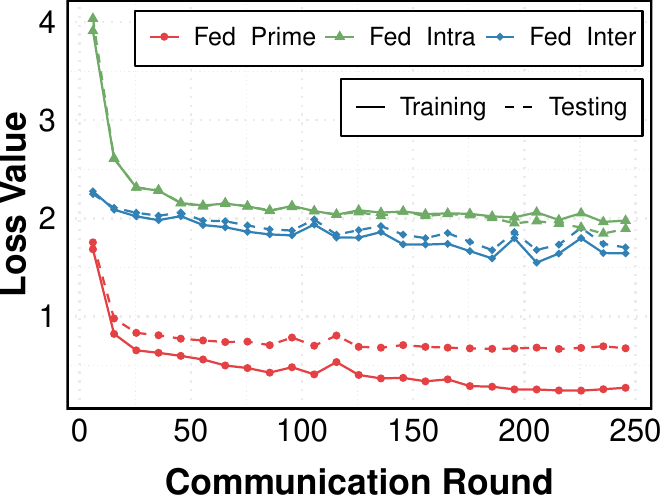}
\caption{Plots of the train/test loss convergence on Food-101 under {\bf Miss-Text}.~\ourmethod~converges fastest.}\vspace{-5mm}
\label{fig:loss_curve}
\end{minipage}
\hfill
\begin{minipage}[t]{0.38\linewidth}
\centering
\includegraphics[height=3.8cm]{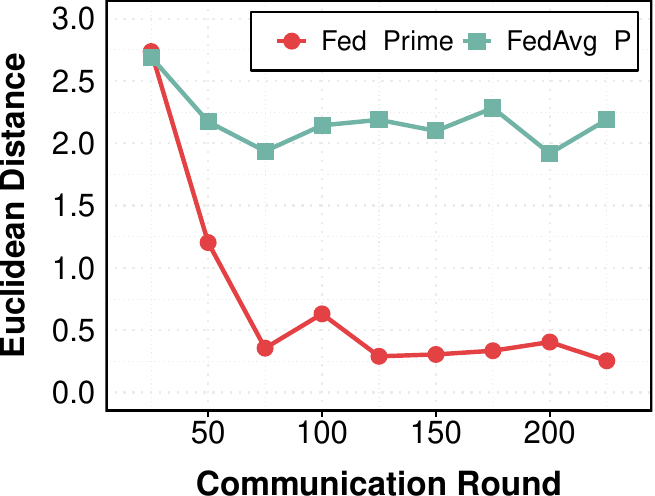}
\caption{~Convergence of Euclidean distance between aggregated prompt centroids on the UPMC Food-101 under {\bf Miss Image} scenario.} \vspace{-8mm}
\label{fig:prompt-distance}
\end{minipage}
\end{figure*}


\subsubsection{Impact of Missing Modalities.}
Across different testing scenarios, we observe that the highest test performance is always achieved when there is no missing data in the test set as expected. Another interesting observation is that in the {\bf Miss Text} training scenario, where approximately $70\%$ of the text features are lost on the training data, the proposed model still performs surprisingly well on the {\bf Text Only} test scenarios. This suggests that the multi-modal prompt alignment is effective in recovering text information, highlighting the proposed design's adaptability. However, in the {\bf Miss Image} training scenario, the test performance on the {\bf Image Only} case suffers instead a significant performance drop. Upon investigation, we believe this is likely due to the VILT backbone’s lightweight image processing module, which was optimized to work in tandem with text for competitive performance, hinting at a potential text-centered nature of VILT during pre-training. This is in fact well-aligned with existing observations that co-training with text often leads to improved performance.

\subsubsection{Robustness Against Various Missing Rates.}
\label{sec:res_robust}
To demonstrate the robustness of \ourmethod~against different missing rates $\eta$, we run multiple comparisons of \ourmethod~, \texttt{FEDAVG-P}, and its centralized variant~\cite{lee2023cvpr} across a wide range of $\eta \in \{0.00, 0.25, 0.50, 0.75, 1.00\}$ under the {\bf Miss Both} training scenarios. 
The reported results in Fig.~\ref{fig:motivation} and Fig.~\ref{fig:miss_both_mean} show that our model significantly outperforms \texttt{FEDAVG-P} and its centralized counterpart \texttt{Centralized-P}. As the missing rate increases, our approach consistently maintains a competitive accuracy above $80\%$, closely approaching the upper‑bound performance of \texttt{FEDAVG-P} in the centralized setting, which benefits from a centralized dataset with better-aligned modalities. These results demonstrate the robustness of our model across varying missing rates and highlight the effectiveness of our federated prompt‑tuning approach, achieving performance comparable to a centralized prompt‑tuning model despite having access only to decentralized local datasets with heterogeneous missing patterns.


\subsubsection{Ablation Study}
\label{sec:res_ablated}
As explained previously in Section~\ref{sec:setting}, \texttt{FED-INTER} and \texttt{FED-INTRA} are in variants of \ourmethod~when we drop the intra- and inter-client prompt sets from the client design, respectively.~Table~\ref{tab:result-food101} shows that \ourmethod~consistently outperforms both \texttt{FED-INTER} and \texttt{FED-INTRA} in all test conditions with substantial gap. The lack of intra-client prompts causes \texttt{FED-INTER} to underperform \ourmethod~suggests that having a dedicated prompt set to capture the common and highly generalizable input-agnostic aspect of fine-tuning is essential to improve performance. Likewise, \texttt{FED-INTRA}, which omits the inter-client prompt set, also shows a significant performance drop. This highlights the importance of capturing and correctly aggregating input-level patterns of missing data through prompt alignment. 



\subsubsection{Convergence Analysis}


\noindent \textbf{Model Convergence.}~Fig.~\ref{fig:loss_curve} shows that \ourmethod~has superior performance, achieving faster convergence and significantly lower final loss values in both training and testing compared to \texttt{FED-INTER} and \texttt{FED-INTRA}. This indicates better learning efficiency and robustness.~Both \texttt{FED-INTER} and \texttt{FED-INTRA} converge more slowly, with consistently higher loss values, especially in the early rounds. Additionally, our method is more stable across server-aggregation rounds, with minimal fluctuations. In contrast, \texttt{FED-INTER} and \texttt{FED-INTRA} exhibit significantly more variability, particularly toward the end. This highlights the effectiveness of \ourmethod, with faster convergence and improved stability across rounds.\vspace{1mm}

\noindent \textbf{Prompt Convergence.}~\cref{fig:prompt-distance} shows that \ourmethod~reduces the distance between the aggregated prompt centroids much faster than \texttt{FEDAVG-P}, which highlights the faster convergence rate of \ourmethod~over \texttt{FEDAVG-P}, indicating faster convergence and better prompt alignment. In contrast, \texttt{FEDAVG-P} converges more slowly and unstably, likely due to incorrect default prompt alignment. This again underscores the necessity of having a learnable prompt clustering mechanism to prevent incorrect prompt alignment. \vspace{1mm}

\section{Conclusion}
\label{sec:conclusion}
Fine-tuning large pre-trained models for specific tasks has become an increasingly effective solution paradigm in machine learning.~However, existing fine-tuning frameworks struggle in practical, multimodal settings where data is distributed across private silos with heterogeneous patterns of missing modalities and features.~To address these challenges, this paper introduces a novel federated prompt-tuning approach that optimizes, aligns, and aggregates prompt instructions, compensating for gaps in data modalities and distributional patterns across a number of diverse and private local datasets.~Rigorous evaluation on multimodal benchmark datasets demonstrates the effectiveness and robustness of our approach against various patterns of missing data in local datasets.~We view this work as an initial step toward developing resilient, adaptable federated models that can fully leverage the diversity of multimodal data in real-world environments. 

\section*{Acknowledgement}
\label{sec:ack}
This research is funded by Hanoi University of Science and Technology (HUST) under grant number T2024-TD-002. Additionally, this work utilized GPU compute resource at SDSC and ACES through allocation CIS230391 from the Advanced Cyberinfrastructure Coordination Ecosystem: Services and Support (ACCESS) program~\cite{ACCESS-resource}, which is supported by U.S. National Science Foundation grants $\#$2138259, $\#$2138286, $\#$2138307, $\#$2137603, and $\#$2138296.
{
    \small
    \bibliographystyle{ieeenat_fullname}
    \bibliography{main}
}

\clearpage
\appendix

\setcounter{page}{1}
\maketitlesupplementary
\renewcommand{\thesection}{\Alph{section}}
\setcounter{section}{0}

\section{Pseudocode for \ourmethod~}
\label{app:a}

\begin{algorithm}[h]
\caption{Client Design}
\label{alg:client}
\textbf{input}: pre-trained model $F$, dataset $D_t$, and $\boldsymbol{w}_g^{inter}$, $\boldsymbol{w}_g^{intra}$ \\
\textbf{output}: intra- and inter-client prompts $\boldsymbol{w}_p^{intra}$, $\boldsymbol{w}_p^{inter}$ 
\begin{algorithmic}[1]
\State initialize prediction head $\boldsymbol{w}_c$
\State initialize $\boldsymbol{w}_p^{intra}$, $\boldsymbol{w}_p^{inter} \leftarrow \boldsymbol{w}_g^{intra}, \boldsymbol{w}_g^{inter}$  
\State optimize $\boldsymbol{w}_c$, $\boldsymbol{w}_p^{intra}$, $\boldsymbol{w}_p^{inter}$ using Eq.~\eqref{eq:5} and Eq.~\eqref{eq:6}
\State\textbf{return} $\boldsymbol{w}_p^{intra}$, $\boldsymbol{w}_p^{inter}$ 
\end{algorithmic}
\end{algorithm}

\begin{algorithm}[h]
\caption{Server Aggregation}
\label{alg:agg}
\textbf{input}: a set of inter-client prompts $(\boldsymbol{w}_t^{inter} \triangleq (\boldsymbol{p}_t^i)_{i=1}^\tau)_{t=1}^n$\\
\textbf{output}: optimized set of aggregated prompts $\boldsymbol{w}_g^{inter}$
\begin{algorithmic}[1]
\State initialize alignment model parameters $\boldsymbol{\theta}$ 
\For{$e=1$ {\bfseries to} $\mathrm{max\text{-}iteration}$}
\State fixing $\boldsymbol{\alpha}$, optimizing $\zeta, \gamma, \boldsymbol{\theta}$ via Eq.~\eqref{eq:11}
\State freezing $\zeta, \gamma, \boldsymbol{\theta}$
\For{$t=1$ {\bfseries to} $n$}       
\State freezing $\boldsymbol{\alpha}_{-t}$
\State optimizing $\boldsymbol{\alpha}_t$ via Eq.~\eqref{eq:12} 
\EndFor
\EndFor
\State $\boldsymbol{w}_g^{inter} \leftarrow \boldsymbol{\theta}$
\For {$q=1$ {\bfseries to} $n\times\tau$}
\State $\boldsymbol{w}_g^{inter} \leftarrow \boldsymbol{w}_g^{inter} \setminus \boldsymbol{\theta}_q$ if $\nexists (t, p): \alpha_t^{p, q} > 0$
\EndFor
\State\textbf{return} set of aggregated inter-client prompts $\boldsymbol{w}_g^{inter}$ 
\end{algorithmic}
\end{algorithm}

\begin{algorithm}[h]
\caption{Multimodal Federated Prompt-Tuning}
\label{alg:FPT}
\textbf{input}: pre-trained model $F$ and no. of iteration $T$\\
\textbf{output}: optimized set of aggregated prompts $\boldsymbol{\theta}$
\begin{algorithmic}[1]
\State initialize global prompt set $\boldsymbol{\theta}$
\For{$e=1$ {\bfseries to} $T$}
\For{$t=1$ {\bfseries to} $n$}
\State send $\boldsymbol{\theta}$ to client $t$
\State request $\boldsymbol{w}_t^{inter}$, $\boldsymbol{w}_t^{intra}$ from client $t$ via Alg.~\ref{alg:client}
\EndFor
\State compute $\boldsymbol{w}_g^{intra}$ run \textsc{FedAvg} on $(\boldsymbol{w}_t^{intra})_{t=1}^n$
\State update $\boldsymbol{w}_g^{inter}$ via running Alg.~\ref{alg:agg} on $(\boldsymbol{w}_t^{inter})_{t=1}^n$
\EndFor
\State\textbf{return} aggregated prompts $\boldsymbol{w}_g^{inter}$, $\boldsymbol{w}_g^{intra}$  
\end{algorithmic}
\end{algorithm}

\begin{figure*}[t]
\centering
\begin{minipage}[t]{0.49\linewidth}
\centering
\includegraphics[height=4.1cm]{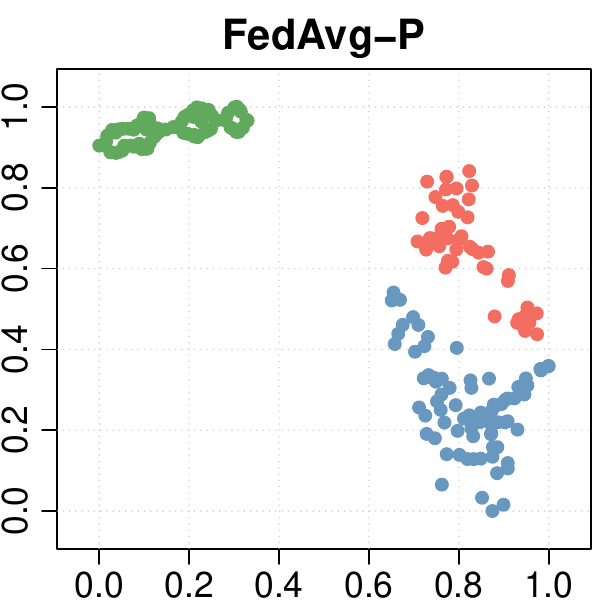}
\hfill
\includegraphics[height=4.1cm]{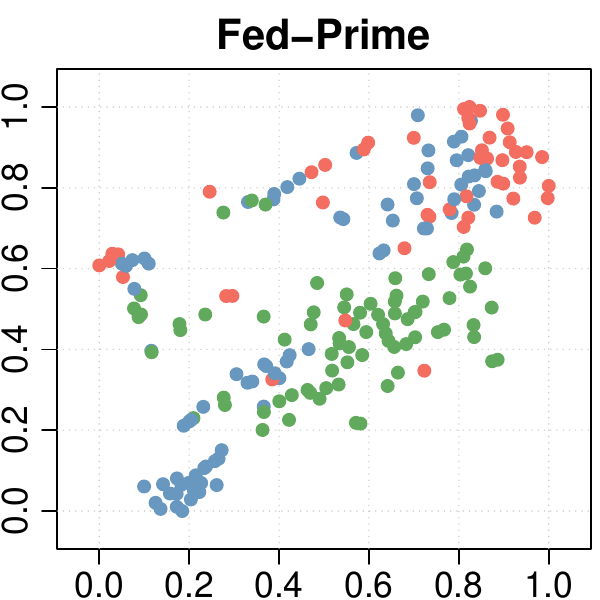}
\end{minipage}
\hfill
\begin{minipage}[t]{0.49\linewidth}
\centering
\includegraphics[height=4.1cm]{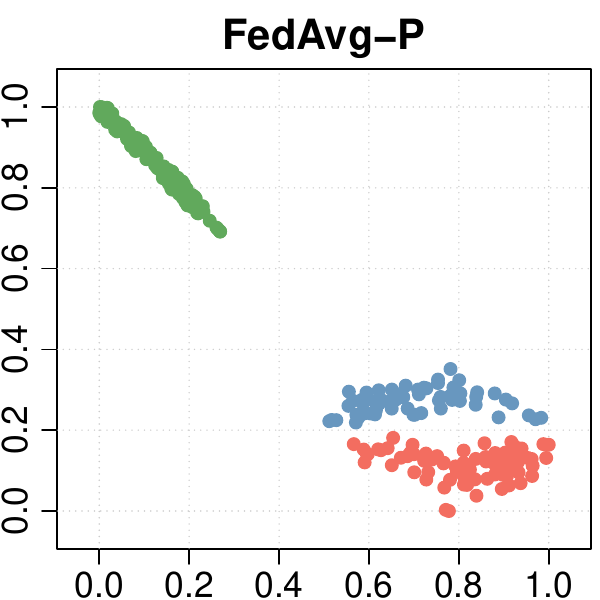}
\hfill
\includegraphics[height=4.1cm]{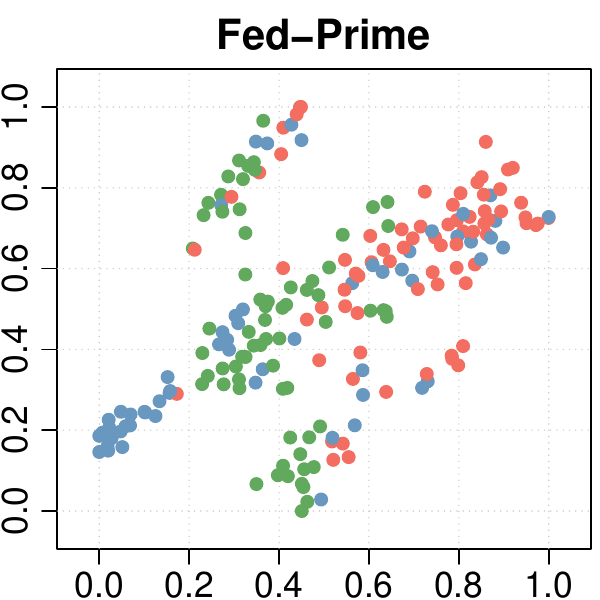}
\end{minipage}

\vspace{0.5em} 
\includegraphics[height=0.7cm]{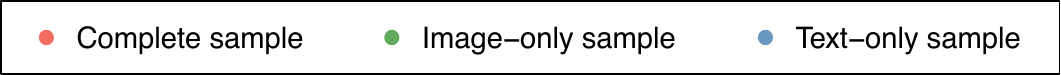}

\caption{t-SNE plots of embeddings prior to classification on MM-IMDB under the \textbf{Miss Both} training scenario for Client \#4 (left) and Client \#14 (right), with two subfigures per client.}
\label{fig:embedding_imdb_clients}
\end{figure*}

\begin{figure*}[h]
\centering
\begin{minipage}[t]{0.24\linewidth}
\centering
\includegraphics[height=4.1cm]{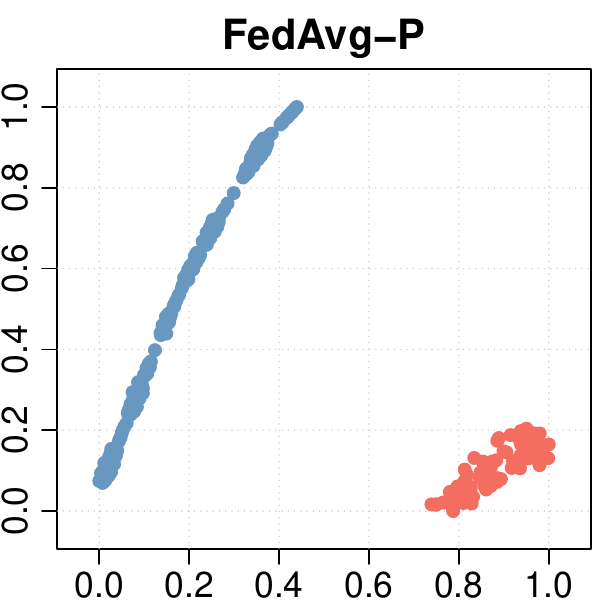}
\end{minipage}
\hfill
\begin{minipage}[t]{0.24\linewidth}
\centering
\includegraphics[height=4.1cm]{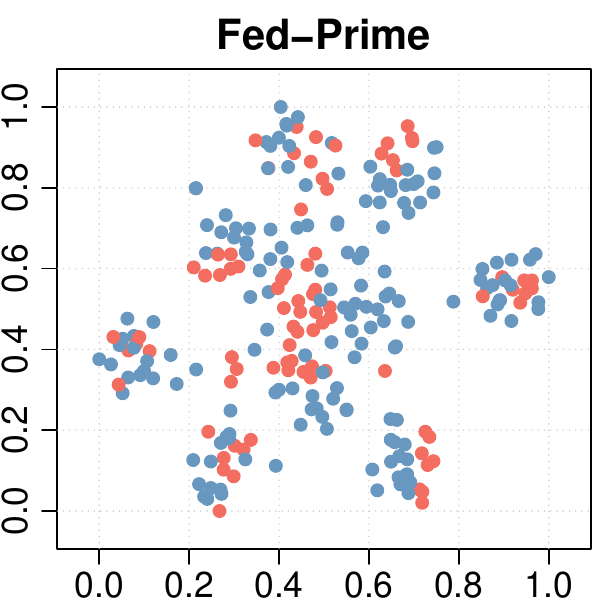}
\end{minipage}
\hfill
\begin{minipage}[t]{0.24\linewidth}
\centering
\includegraphics[height=4.1cm]{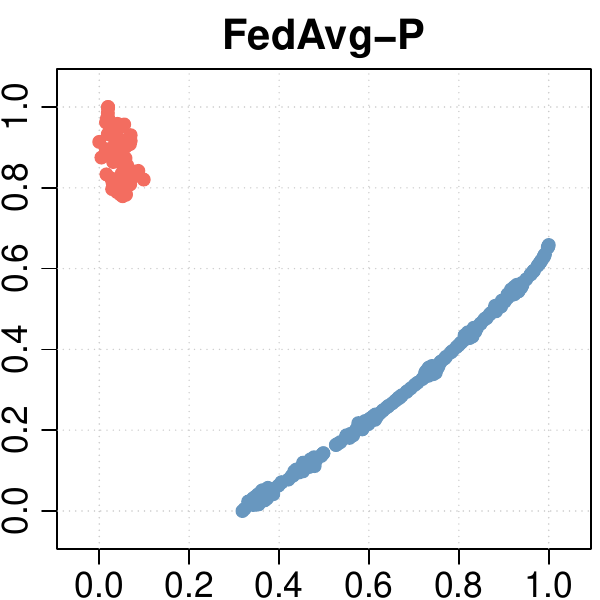}
\end{minipage}
\hfill
\begin{minipage}[t]{0.24\linewidth}
\centering
\includegraphics[height=4.1cm]{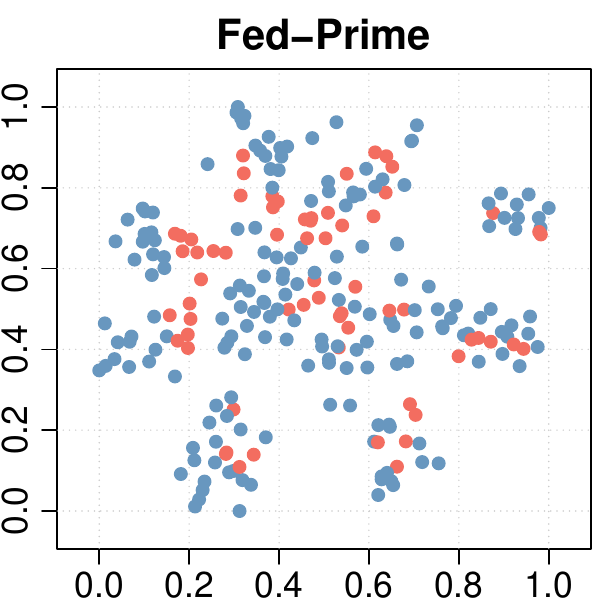}
\end{minipage}

\vspace{0.5em} 
\includegraphics[height=0.7cm]{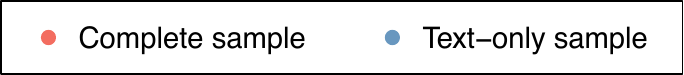}

\caption{t-SNE plots of embeddings prior to classification on UMPC Food-101 under the \textbf{Miss Image} training scenario for Client \#4 (left) and Client \#14 (right), with two subfigures per client.}
\label{fig:embedding_food_clients}
\end{figure*}

\section{Implementation Details}
\label{app:b}
\noindent \textbf{Input.} 
All baselines use inputs from the UPMC Food-101 dataset, comprising $6,728$ multimodal samples, and the MM-IMDB dataset, which includes $5,778$ image-text pairs after the preprocessing. For the text modality, inputs are tokenized using the BERT-base-uncased tokenizer, as outlined in \cite{lee2023cvpr}, with a maximum sequence length of $40$ for UPMC Food-101 and $128$ for MM-IMDB.
When text is missing, we use an empty string as a dummy input. For the image modality, we follow \cite{lee2023cvpr, kim2021vilt} by resizing the shorter side of the input image to $384$ pixels, while keeping the longer side under $640$ pixels to maintain the aspect ratio. As in \cite{kim2021vilt}, we decompose images into \(32 \times 32\) patches. If the image is missing, we create a dummy image with all pixel values set to one, as described in \cite{lee2023cvpr}.

\noindent \textbf{Multimodal Backbone.} Following~\cite{lee2023cvpr}, we adopt the pretrained multimodal transformer ViLT~\cite{kim2021vilt} as our backbone as it is commonly used in various transformer-based methods for learning multimodal tasks. ViLT stems from Vision Transformer~\cite{dosovitskiy2020vit} and advances to process multimodal inputs with tokenized texts and patched images. 
Without using modality-specific feature extractors, ViLT is pretrained on several large vision-language datasets (e.g., MS-COCO~\cite{lin2014mscoco} and Visual Genome~\cite{Krishna2016VisualGC}) via objectives such as Image Text Matching (ITM) and Masked Language Modeling (MTM).

\noindent \textbf{Model Training Details.}
To reduce the need for extensive fine-tuning, we freeze the ViLT backbone parameters and train only the learnable prompts and task-specific layers (pooler and classifier). 
Each pool consists of $20$ prompts, from which $5$ prompts are selected per input from each pool—inter- and intra-client ones. These prompts are concatenated and added to the initial Multi-Head Self-Attention (MSA) layer, resulting in a total of $10$ prompts for each input.
For \texttt{FED-INTRA} and \texttt{FED-INTER}, where only a single pool is utilized, the pool still contains $20$ prompts. From this pool, $10$ prompts are directly selected, ensuring that the total number of prompts per input remains consistent with \ourmethod.
In contrast, \texttt{FEDAVG-P} and \texttt{FEDMSPLIT-P} attach prompts to the first $6$ MSA layers, with each prompt having a length of $16$, resulting in a greater number of prompts per input.

\noindent \textbf{Baseline Aggregation Details.} 
In \texttt{FEDAVG-P}, the server aggregation procedure updates the newly trained components—namely, the prompts, pooler, and classifier—for each modality set and subsequently distributes them to the respective clients.
In \texttt{FEDMSPLIT-P}, the original work assumes that each client possesses a distinct set of modalities, with a uniform modality distribution across all samples within a given client. This assumption contrasts with our setting, where clients exhibit heterogeneous and incomplete multimodal data distributions. Furthermore, the original framework constructs modality-specific encoder blocks, aggregating similar blocks based on similarity coefficients that quantify inter-client modality relationships. To adapt this approach to our setting, we aggregate all modular components—including prompts, pooler, and classifier—based on client similarity while assuming that all clients maintain a shared set of these modular components, encompassing all possible modality configurations.

\begin{table}[t]
\caption{Top-1 accuracy of \ourmethod~across varying pool sizes, reported on UPMC Food-101 dataset.}
\label{tab:vary_pool_size}
\centering
\resizebox{0.9\linewidth}{!}{%
\begin{tabular}{@{}l|ccccc@{}}
\toprule
\textbf{Pool Size}           & 10    & 15    & 20    & 25    & 30    \\ \midrule
\textbf{Top-1 Test Accuracy} & 73.94 & 82.50 & 82.61 & 82.72 & 83.21 \\ \bottomrule
\end{tabular}%
}
\end{table}

\noindent \textbf{Hyperparameter Settings.}
All evaluated baselines utilize a batch size of $512$ for the UPMC Food-101 dataset and $256$ for MM-IMDB. 
The training datasets for UPMC Food-101 and MM-IMDB were randomly partitioned into $20$ clients, all of whom participated in every communication round.
We use a Stochastic Gradient Descent (SGD) optimizer with a base learning rate of $0.05$ for Food-$101$ and $0.01$ for IMDB, with a communication round of $250$ for faster convergence compared to full fine-tuning.
During each round, clients train the global model locally for $1$ epoch.
The hyperparameter configurations were consistently applied across experimental missing scenarios. 

\noindent \textbf{Code Availability and Reproducibility.}
We release our full implementation and configurations at \url{https://github.com/hangpt01/FedPrime}. 
The repository includes source code, configuration files (e.g., prompt dimensions, number of prompts, clustering thresholds, regularization coefficients), and brief guidelines with scripts to run experiments or adapt the framework to other datasets.

\noindent \textbf{Automatic Hyperparameter Tuning.}
Beyond the default settings described above, the repository includes utilities and scripts to automatically tune key hyperparameters (e.g., clustering iteration thresholds, pool sizes, and the number of active prompts). 
We also provide implementations of common tuning methods, namely Grid Search, Random Search, and Bayesian Optimization, which allow users to efficiently explore configurations based on validation loss or heuristic rules. 
These tools reduce the need for extensive manual searches and simplify adaptation to new problems.

\begin{table}[t]
\centering
\caption{Resource required of different baselines in each round}
\label{tab:resource}
\resizebox{\linewidth}{!}{%
\begin{tabular}{@{}l|rr|rr@{}}
\toprule
\multirow{2}{*}{\textbf{Method}} & \multicolumn{2}{c|}{\textbf{UPMC Food-101~\cite{upmcfood101}}}                        & \multicolumn{2}{c}{\textbf{MM-IMDB~\cite{arevalo2017IMDB}}}                        \\ \cmidrule(l){2-5} 
                                 & \multicolumn{1}{c}{GPU (GB)} & \multicolumn{1}{c|}{Time (s)} & \multicolumn{1}{c}{GPU (GB)} & \multicolumn{1}{c}{Time (s)} \\ \midrule
\texttt{FEDAVG-P}           & 31.70 & 245.53 & 39.07 & 649.58 \\
\texttt{FEDMSPLIT-P}         & 34.30 & 265.19 & 39.13 & 755.62 \\
\texttt{FED-INTER}          & 16.83 & 315.67 & 23.73 & 685.27 \\
\texttt{FED-INTRA}          & 16.65 & 253.60 & 23.31 & 625.83 \\ \midrule
\textbf{\ourmethod} & 16.80 & 240.54 & 23.68 & 653.57 \\ \bottomrule
\end{tabular}%
}
\end{table}

inst

\section{Computational Resource Analysis.}
Table~\ref{tab:resource} summarizes the computational resources required by the evaluated methods. Notably, \texttt{FEDAVG-P} and \texttt{FEDMSPLIT-P} demonstrate substantially higher GPU memory usage, such as $31.70$ GB and $34.30$ GB on the UPMC Food-101 dataset, nearly double that of \ourmethod. This discrepancy can be attributed to their strategy of prepending prompts in the first six Multi-Head Self-Attention (MSA) layers out of a total of $12$ layers, which expands sequence lengths and amplifies memory demands for intermediate activations, gradients, and computations. Moreover, \texttt{FEDMSPLIT-P} further increases computational costs, as it requires estimating similarity across clients and maintaining a personalized model for each client. Conversely, \ourmethod~limits the prompt addition to only the first MSA layer, thereby restricting the augmented sequence length and associated computations to a single layer. This design significantly reduces the overall memory overhead. Other variants demonstrate relatively comparable GPU memory usage.
Overall, GPU usage for the MM-IMDB dataset is higher than for UPMC Food-101, primarily due to the longer text sequence lengths employed in the experiments.

\noindent In terms of execution time per round, \texttt{FED-INTER} and \texttt{FEDMSPLIT-P} interchangeably exhibits the longest runtime.
Other methods exhibit similar runtime performance, with approximately a $5$-second difference between \texttt{FEDAVG-P} and \ourmethod. Importantly, \ourmethod~demonstrates its efficiency by achieving high performance while maintaining lower GPU memory requirements and comparable execution time.

\section{Additional Ablation Studies}
Table~\ref{tab:vary_pool_size} demonstrates a clear positive correlation between pool size and test accuracy, with larger pool sizes consistently yielding higher accuracy. A significant drop in accuracy is observed when the pool size decreases from $15$ to $10$, highlighting a critical threshold for maintaining performance. While increasing the pool size from $15$ to $30$ results in an overall improvement in accuracy, the rate of improvement diminishes as the pool size grows from $15$ to $25$. Notably, a pool size of $30$ achieves the highest accuracy, showing a larger improvement compared to the smaller gains observed when increasing the pool size from $15$ to $25$. 
In general, while larger pool sizes improve accuracy, the marginal gains may not justify the extra computational cost. Hence, a pool size of $20$ is selected for all subsequent experiments in this study, as it provides a practical balance between accuracy and computational efficiency.

\section{Additional Prompting Analysis}
\cref{fig:embedding_imdb_clients} illustrates the embeddings at the final round (round 250) for \texttt{FEDAVG-P} and \ourmethod~under the \textbf{Miss Both} scenario on the MM-IMDB dataset, visualized for two random clients (Client \#4 and Client \#14). As shown in the figure, \texttt{FEDAVG-P},  employing a design that prompts embeddings for each missing type separately, possesses distinct clusters for each sample type in the embedding space. In contrast, \ourmethod~produces more scattered embeddings, where the embeddings of complete samples are positioned closer to those of samples with missing modalities. This demonstrates \ourmethod's superior ability to learn meaningful and comprehensive representations, regardless of the specific missing type in a sample.
A notable observation is that image-only samples in \texttt{FEDAVG-P} embeddings are significantly distant from text-only and complete samples. This separation highlights the limitations of \texttt{FEDAVG-P} in capturing cross-modal relationships effectively, which can be attributed to the fact that this method utilizes prompts specific to the missing modality type. This further supports the hypothesis that the ViLT backbone in \ourmethod~has a better capability for text representation and integration.

\noindent To further explore these models, \cref{fig:embedding_food_clients} presents embeddings under the \textbf{Miss Image} training scenario on the UMPC Food-101 dataset for the same clients. A similar trend is observed: \ourmethod~generates embeddings with well-integrated representation of different sample types, while \texttt{FEDAVG-P} produces two distinct and distant clusters. 
As observed before, when image-only samples are included in the data, embeddings for text and complete samples in \texttt{FEDAVG-P} may cluster closer together. However, even in the absence of image-only samples, embeddings for text and complete samples remain overly distinct, highlighting a lack of cohesive representation across modalities. This separation likely hinders \texttt{FEDAVG-P}'s performance, contributing to its moderate results when classifiers are subsequently applied to these embeddings.




\section{Additional Experiment Results}

\begin{figure*}[t]
\centering
\begin{minipage}[t]{0.32\linewidth}
\centering
\includegraphics[height=4.8cm]{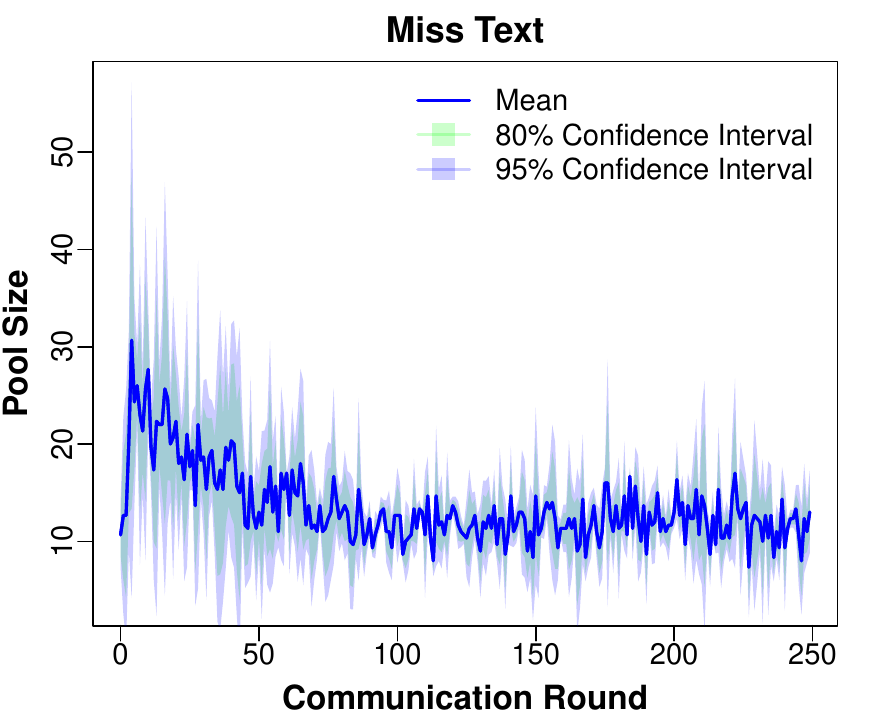}
\end{minipage}
\hfill
\begin{minipage}[t]{0.32\linewidth}
\centering
\includegraphics[height=4.8cm]{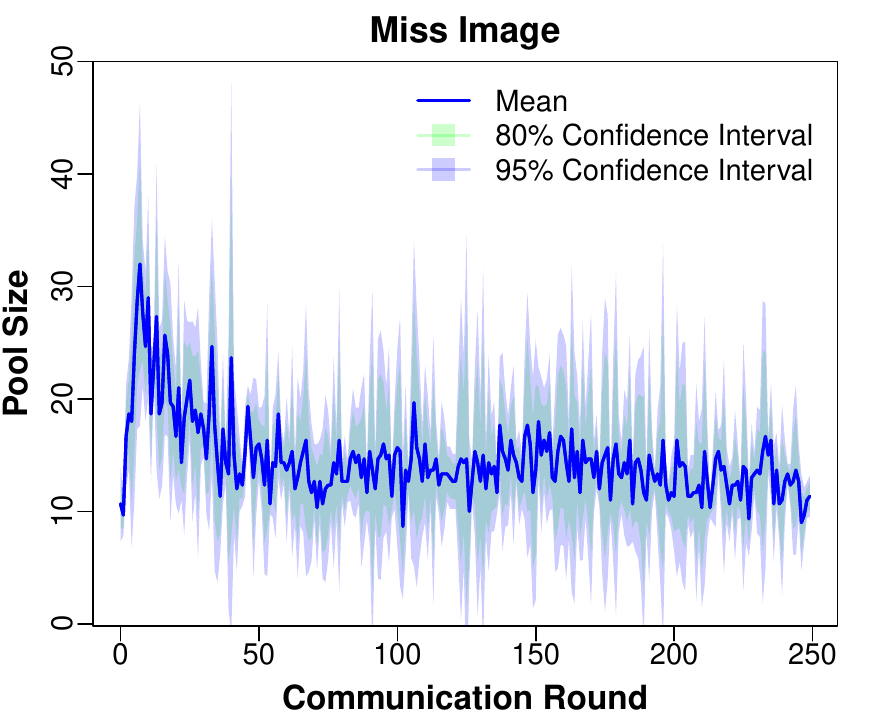}
\end{minipage}
\hfill
\begin{minipage}[t]{0.32\linewidth}
\centering
\includegraphics[height=4.8cm]{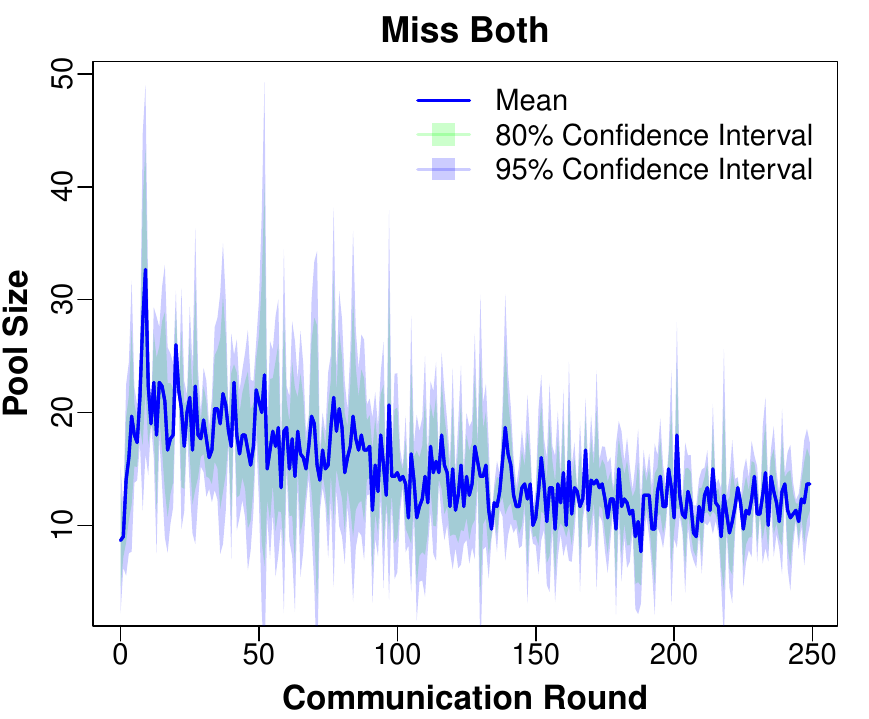}
\end{minipage}

\caption{Variations in UMPC Food-101 inter-client prompt pool size across $250$ communication rounds under different training scenarios.}
\label{fig:food_pool_size}

\end{figure*}

\begin{figure*}[t]
\centering
\begin{minipage}[t]{0.32\linewidth}
\centering
\includegraphics[height=4.8cm]{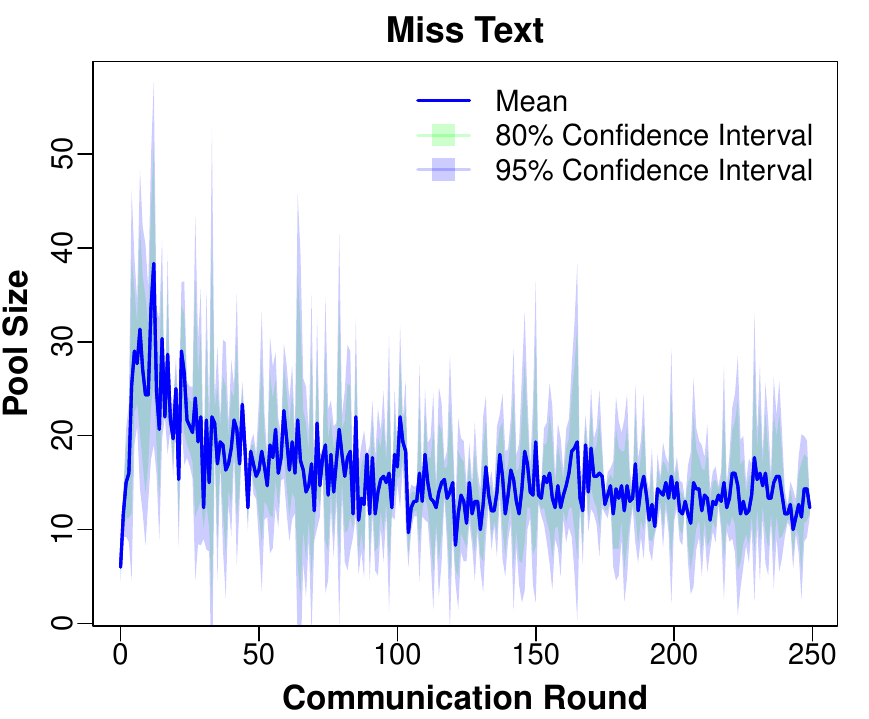}
\end{minipage}
\hfill
\begin{minipage}[t]{0.32\linewidth}
\centering
\includegraphics[height=4.8cm]{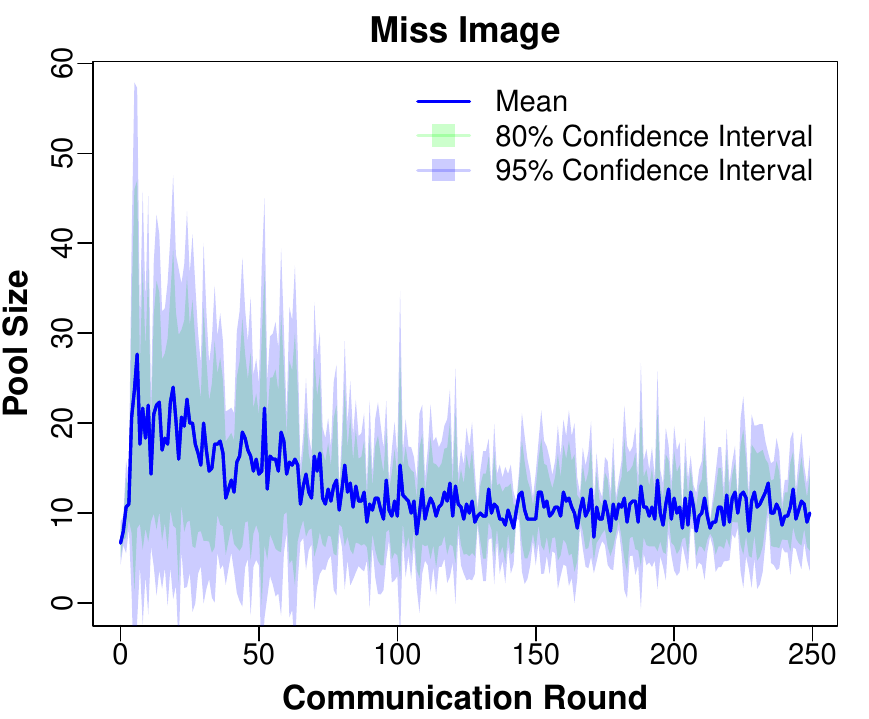}
\end{minipage}
\hfill
\begin{minipage}[t]{0.32\linewidth}
\centering
\includegraphics[height=4.8cm]{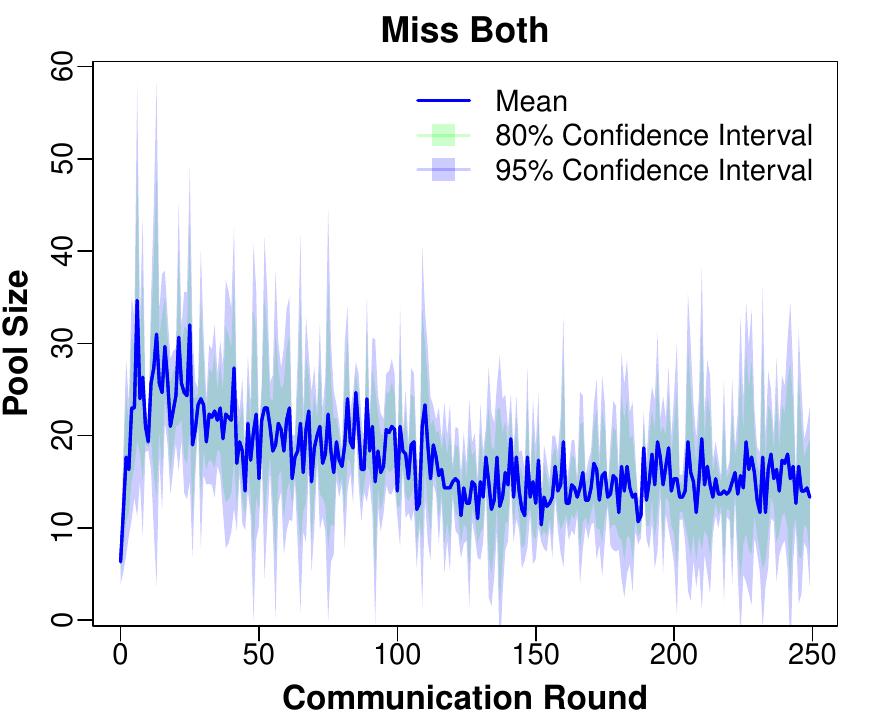}
\end{minipage}

\caption{Variations in MM-IMDB inter-client prompt pool size across $250$ communication rounds under different training scenarios.}
\label{fig:imdb_pool_size}

\end{figure*}

\textbf{Inter-pool Convergence.}
\cref{fig:food_pool_size} and \cref{fig:imdb_pool_size} depict the size of the pool of the inter-client prompt pool for the two datasets in three defined training scenarios. The results indicate an initial increase in the spread of prompts relative to their centroid, followed by a subsequent decrease. This observation suggests that the proposed method optimizes the learned prompts to effectively capture the diversity of data across participating clients. Once the model sufficiently learns the clustering and alignment, it begins to condense client-specific information, which leads to an increase in the overall pool size.

\noindent \textbf{Scalability in Large-scale FL.}
We performed additional experiments on the Food-101 dataset, simulating a scenario with 100 clients and varying client participation rates between 10\% and 50\%. These results were compared against the strongest baseline for this dataset, \texttt{FEDMSPLIT-P}. As shown in Tab.~\ref{tab:result-large_scale}, our dual-prompt mechanism demonstrates strong scalability, with performance improving as the number of participating clients increases. Moreover, our method consistently outperforms the baseline in terms of test accuracy.
In terms of computational efficiency, \texttt{FED-PRIME} exhibits substantially lower GPU memory consumption, which remains stable regardless of the number of clients, in contrast to \texttt{FEDMSPLIT-P}, which experiences increased memory usage as the number of clients increases. This property makes \texttt{FED-PRIME} particularly well-suited for deployment in resource-constrained environments.

\noindent \textbf{Performance under Highly Imbalanced Data.}
In addition to modality-missing heterogeneity, we also evaluate our method under extreme data imbalance. 
FedProx~\cite{li2020federated}, a popular baseline specifically designed to address Non-IID problems, is selected for comparison in this setting. 
Specifically, we compare our method against a variant that modifies the averaging of classifiers and intra-client prompts using FedProx, denoted as \texttt{FEDPROX-P}. 
We conduct additional evaluations on the Food-101 dataset under extreme Non-IID settings by simulating a Dirichlet distribution with $\alpha = 0.1$. 
The results from Tab.~\ref{tab:result-noniid} confirm that our method consistently outperforms \texttt{FEDPROX-P}, further demonstrating the effectiveness of aggregating inter-client prompts to combat extreme data imbalance. 
Although \texttt{FEDPROX-P} is designed to address data heterogeneity, our method’s use of inter-client prompts effectively mitigates this issue, highlighting the strength of our prompt-alignment algorithms and the selective averaging strategy applied to specific model components. 
We also evaluate scenarios in which clients possess either full modalities or only a single modality, with our method surpassing both \texttt{FEDAVG-P} and its centralized counterpart in accuracy (see Fig.~\ref{fig:motivation} and Fig.~\ref{fig:miss_both_mean}).

\begin{table*}[htb]
\vspace{-8pt}
    \centering
    \caption{Non-IID FL settings with UPMC Food-101 Dataset. 
    Results indicated by a dash (-) represent scenarios where \textbf{Test (Miss Both)} is the same as \textbf{Test ($\sim$Train)}.
    \label{tab:result-noniid}}
    \resizebox{0.72\linewidth}{!}{%
    \begin{tabular}{@{}c|l|rrrrr@{}}
    \toprule
    \multicolumn{1}{l|}{\textbf{Train}} &
      \textbf{Method} &
      \multicolumn{1}{c}{\textbf{\begin{tabular}[c]{@{}c@{}}Test\\ ($\sim$ Train)\end{tabular}}} &
      \multicolumn{1}{c}{\textbf{\begin{tabular}[c]{@{}c@{}}Test\\ (Miss Both)\end{tabular}}} &
      \multicolumn{1}{c}{\textbf{\begin{tabular}[c]{@{}c@{}}Test\\ (Full Modal)\end{tabular}}} &
      \multicolumn{1}{c}{\textbf{\begin{tabular}[c]{@{}c@{}}Test\\ (Text only)\end{tabular}}} &
      \multicolumn{1}{c}{\textbf{\begin{tabular}[c]{@{}c@{}}Test\\ (Image only)\end{tabular}}} \\ \midrule
    \multirow{2}{*}{\begin{tabular}[c]{@{}c@{}}Miss\\ Text\end{tabular}} &
      \texttt{FEDPROX-P} &
      67.42 &
      64.34 &
      77.29 &
      56.83 &
      68.24 \\
     &
      \textbf{\texttt{FED-PRIME}} &
      \textbf{71.20} &
      \textbf{71.15} &
      \textbf{85.08} &
      \textbf{63.91} &
      \textbf{69.56} \\ \midrule
    \multirow{2}{*}{\begin{tabular}[c]{@{}c@{}}Miss \\ Image\end{tabular}} &
      \texttt{FEDPROX-P} &
      82.56 &
      71.26 &
      85.24 &
      83.05 &
      45.31 \\
     &
      \textbf{\texttt{FED-PRIME}} &
      \textbf{87.38} &
      \textbf{75.59} &
      \textbf{89.25} &
      \textbf{87.05} &
      \textbf{48.47} \\ \midrule
    \multirow{2}{*}{\begin{tabular}[c]{@{}c@{}}Miss \\ Both\end{tabular}} &
      \texttt{FEDPROX-P} &
      75.75 &
      - &
      89.36 &
      83.98 &
      69.61 \\
     &
      \textbf{\texttt{FED-PRIME}} &
      \textbf{79.98} &
      - &
      \textbf{91.00} &
      \textbf{86.70} &
      \textbf{70.38} \\ \bottomrule
    \end{tabular}%
    }
    \vspace{-10pt}
\end{table*}

\end{document}